\documentclass[pra,aps,amsmath,amssymb,twocolumn,floatfix,superscriptaddress]{revtex4-1}
\usepackage{amsmath}
\usepackage{graphicx}
\usepackage{color}
\usepackage{amssymb}
\usepackage[utf8]{inputenc}
\usepackage[T1]{fontenc}
\usepackage{hyperref}

\begin{document}
\title{Effect of Rashba spin-orbit and Rabi couplings on the excitation spectrum of binary Bose-Einstein condensates}

\author{Rajamanickam Ravisankar}
\affiliation{Department of Physics, Indian Institute of Technology, Guwahati 781039, Assam, India} 
\affiliation{Department of Physics, Bharathidasan University, Tiruchirappalli 620024, Tamilnadu, India}
\author{Henrique Fabrelli}
\affiliation{Instituto de F\'{i}sica, Universidade de S\~{a}o Paulo, 05508-090 S\~{a}o Paulo, Brazil}
\author{Arnaldo Gammal}
\affiliation{Instituto de F\'{i}sica, Universidade de S\~{a}o Paulo, 05508-090 S\~{a}o Paulo, Brazil}
\author{Paulsamy Muruganandam}
\affiliation{Department of Physics, Bharathidasan University, Tiruchirappalli 620024, Tamilnadu, India}
\author{Pankaj Kumar Mishra}
\affiliation{Department of Physics, Indian Institute of Technology, Guwahati 781039, Assam, India}

\date{\today}

\begin{abstract} 
We present the collective excitation spectrum analysis of binary Bose-Einstein condensates (BECs) with spin-orbit (SO) and Rabi couplings in a quasi-two-dimensional system. In particular, we investigate the role of SO and Rabi coupling strengths in determining the dynamical stability of the coupled BECs using Bogoliubov-de Gennes (BdG) theory. Using the eigenergy of BdG spectrum, we confirm the existence of phonon, roton, and maxon modes with weak repulsive intra- and inter-species contact interactions.  The depth of the minimum corresponding to the roton mode depends strongly on the coupling strength. We find that the increase of the SO coupling leads to instability, while the increase in the Rabi coupling stabilizes the system. Also the eigenvectors of BdG spectrum indicates the presence of density like mode in the stable regime and spin like  modes in unstable regimes.  A phase diagram demonstrating the stability regime in the plane of SO and Rabi coupling strengths is obtained. Finally, we complement the observation of the excitation spectrum with the direct numerical simulation results of coupled Gross-Pitaevskii equations. 
\end{abstract}

\flushbottom

\maketitle
\section{Introduction}
The experimental realization of Bose-Einstein condensates (BECs) in dilute atomic gases has triggered immense interest in the physics of ultracold matter~\cite{Anderson1995, Davis1995, Bradley1995}. Since then, BECs has become an excellent system for manipulating many of the macroscopic phenomena through the controlled environment in the quantum regime. 
Last few decades have seen an upsurge in the research of BEC in particular understanding the fundamental and dynamical aspects of 
solitons, its behaviour under the  optical lattice and disordered potentials trap, superfluid-Mott insulator phase transition, presence of localization, dipolar and spin-orbit (SO) coupled BECs, etc.~\cite{Gerton2000, Greiner2002, Morsch2006, Lewenstein2007, Roati2008, Chin2010, Griesmaier2005, Lu2011, Aikawa2012, Lin2011, Galitski2013}. In this paper we investigate the effect of the Rabi and Rashba spin-orbit coupling on the dynamical stability of the binary BEC system.  

In BECs, the internal atomic  states can  be manipulated to produce quite novel systems like binary and multi-component condensates. These binary or multi-component BECs consist of different isotopes or hyperfine states of the same or the different atomic species, which are coherently coupled by the external fields~\cite{Ho1998, Papp2008}. This facilitates the testing ground of plethora of macroscopic quantum many-body phenomena such as quantum turbulence~\cite{Takeuchi2010}, quantum phase transitions~\cite{Sabbatini2011}, quantized vortices, matter wave solitons~\cite{Law2010}, vortex-antivortex~\cite{Wen2013} etc.  BECs of bosons in two different hyperfine states, designated as spin-$1/2$ bosons, have opened up a new way for synthetic SO coupling, a key ingredient for many important condensed matter phenomena. In this connection, realization of SO coupled Bose and Fermi gases at ultra-low temperature has paved the way for many important physical phenomena of current interest such as measurement of spin Hall effect~\cite{Galitski2013, Aidelsburger2013}, topological insulators~\cite{Goldman2010}, topological superfluids~\cite{Wu2016}, atomtronics (or spintronics)~\cite{Seaman2007}, and quantum computing~\cite{Andrianov2014}.

Numerical simulations have played an important role in unravelling different interesting phases in coupled BECs. The stability and dynamics of matter-wave bright and dark solitons in one-dimensional SO coupled BECs have been investigated quite extensively using coupled Gross-Pitaevskii equations (GPEs)~\cite{Ravisankar2020sol, Achilleos2013-bs, Achilleos2013}. In two dimension Rashba SO coupled BECs with weak harmonic trap exhibits plane and stripe wave phases while under strong harmonic trap it displays the presence of  vortex pairs, honeycomb-lattice and half-quantum vortices upon varying the Rabi (zero momentum) coupling strength~\cite{Jin2014}. The SO coupled BEC's confined in the optical lattice reveal different ground state structures as the form of vortex-antivortex pair~\cite{Li2013}. Using the numerical and variational analysis Cheng \textit{et al.} demonstrated that SO coupled BECs trapped under bichromatic optical lattices show Anderson Localization~\cite{Cheng2014}. It was found that the Rabi coupling stabilizes the superfluid phase in coupled BECs in optical lattices~\cite{He2021}. A variety of collective modes, namely, Nambu-Goldstone, slosh, bifurcation modes were studied in the trapped two-component quasi-two-dimensional (quasi-2D) BECs~\cite{Pal2017, Pal2018}.

The stability of different ground state phases of SO coupled BECs could be well understood by analyzing the spectrum of elementary excitation. For instance, the Bogoliubov-de Gennes (BdG) spectrum, much related to macroscopic quantum phenomena, such as, superfluidity and superconductivity, provide the fundamental information about the condensate dynamics. The excitation spectrum of BECs with Rashba-Dresselhaus SO coupling found to exhibit roton-maxon structures~\cite{Martone2012, Zheng2013, Khamehchi2014, Ji2015}. Using Green's function technique various interesting features, like, multi-criticality, metastability, and the roton were found to exist in three dimensional Rashba SO coupled BEC~\cite{Liao2015}. The metastability could be understood as a result of  the absence of imaginary frequencies in the BdG spectrum.
Ozawa \textit{et al.} numerically investigated the dynamical and energetic instabilities in quasi-1D SO and Raman coupled BECs~\cite{Ozawa2013}. Spin-dipole and breathing modes of the collective excitation spectrum give clear picture of phase boundaries, which was also confirmed by the quenching dynamics numerically~\cite{Chen2017}. The effect of SO, Rabi couplings and nonlinear interactions provide collective oscillations, which has a transition from harmonicity to anharmonicity~\cite{Yu2018}. In a recent work Geier \textit{et al.} found the signature of Goldstone modes in harmonically trapped SO coupled BECs~\cite{Geier2021}.

The application of BdG spectrum  was studied in the context of exploring the superfluid phase  in SO coupled BECs~\cite{Zhu2012}. Further Yu \textit{et al.} used this idea in obtaining the ground state phase diagram particularly, excitations of zero momentum phases in quasi-2D SO coupled BECs~\cite{Yu2013}. In the liquid phase it was demonstrated that the quasi-2D BECs exhibits different excitation modes, like, phonon, roton and double roton modes with different SO couplings as the  interaction strengths are varied~\cite{Sahu2020}. 
 
It may be noted that most of the studies on the collective excitations are mainly focused on quasi-1D spin-orbit coupled BECs~\cite{Martone2012, Zheng2013, Khamehchi2014, Ji2015} and limited explorations are available on the stability of the superfluid phases in two dimensions~\cite{Wang2010}. For obvious reasons, it would be more appropriate to study collective excitations in higher spatial dimensions. In this paper we present a detailed study on the stability analysis of SO coupled BECs in two dimensions from the excitation spectrum of Bogoliubov-de Gennes (BdG) equations. In particular, we carry out a systematic analysis on collective excitations of SO coupled BECs in quasi-two dimensions using the dispersion relations obtained by the application of Bogoliubov theory to the coupled Gross-Pitaevskii equations (GPEs). These dispersion relations are then used for the stability analysis of plane waves, phonon-maxon-roton excitations and the interplay between SO and Rabi couplings. The roton mode is the precursor of the stripe phase with periodic fringes~\cite{Ji2015} and instability is a fundamental ingredient of the existence of matter-wave solitons.

The paper is organized as follows. In Sec.~\ref{sec:2}, we introduce mean-field theoretical model used for the study. In Sec.~\ref{sec:3}, we analytically derive the single particle dispersion relation. Following this, using Bogoliubov-de-Gennes method, we investigate the collective excitation of Rashba SO coupled BECs  analytically and numerically  by computing the eigenspectrum and eigenvectors in Sec.~\ref{sec:4}.  The numerical simulation of the stability analysis of the ground state obtained by solving the coupled GPEs of quasi-two dimensional Rashba SO coupled BEC is presented in Sec.~\ref{sec:5}. 
Finally in Sec.~\ref{sec:7} we conclude our observation of the SO coupled BEC.

\section{Mean-field model of coupled BEC\lowercase{s}}
\label{sec:2}
We consider a pseudospin-$1/2$ Bose-Einstein condensate with Rashba spin-orbit and Rabi coupling which Hamiltonian is given by~\cite{Jin2014}
\begin{align}
\mathcal{H} = \mathcal{H}_{0} + \mathcal{H}_{I}, \label{totHam} \end{align}
with 
\begin{align}
\mathcal{ H}_{0}=& \int \Psi^\dagger\left[\frac{\boldsymbol{p}^2}{2m}+{V} 
+ {k_L'}\boldsymbol{p} \cdot \boldsymbol{\sigma} + {\Omega'} \sigma_{x}\hbar \right] \Psi d \mathbf{r}, \\
\mathcal{H}_{I}=& \int \left[ \frac{ {g}_{\uparrow \uparrow}}{2} \vert \Psi_{\uparrow}\vert^2+\frac{ {g}_{\downarrow \downarrow}}{2} \vert\Psi_{\downarrow}\vert^2 +  {g}_{\uparrow \downarrow}  \vert\Psi_{\uparrow}\vert \vert\Psi_{\downarrow}\vert\right] d\mathbf{r}.
\end{align}
Here $\Psi = \left(\Psi_{\uparrow} \; \Psi_{\downarrow}\right)^T$ is the two component spinor normalized wave functions that satisfy the condition $\int \left( \vert\Psi_{\uparrow}\vert^2 + \vert\Psi_{\downarrow}\vert^2\right) dr = N$, with $N$ being the total number of particles, $m$ is the atomic mass, and $\boldsymbol{p} = -\mathrm{i}\hbar\boldsymbol{\nabla}$ is the momentum operator, $\boldsymbol{\sigma} =(\sigma_x,\sigma_y)$ are the $2\times2$ Pauli matrices, $ {g}_{\uparrow \uparrow}$ and ${g}_{\downarrow \downarrow} = 4\pi  a \hbar^2/m$, $(a = a_{\uparrow \uparrow}= a_{\downarrow \downarrow})$, $ {g}_{\uparrow \downarrow}= 4\pi  a_{\uparrow \downarrow} \hbar^2/m$, are the intra- and inter-species contact interaction strengths with $a$ and $a_{\uparrow \downarrow}$ being the intra- and inter-species $s$-wave scattering lengths, respectively. We consider the condensates confined in the harmonic trap potential with form $V= m[ \lambda^2 x^2 + \kappa^2 y^2 +\eta^2 z^2]/2$,  where the trap aspect ratios are $\lambda = \omega_{x}/\omega_{\perp}$, $\kappa = \omega_{y}/\omega_{\perp}$ and $\eta = \omega_{z}/\omega_{\perp} \gg 1$.

In order to study the ground state and dynamical properties of Rashba SO coupled BECs with strong axial traps, we consider the two-dimensional coupled GPEs in dimensionless form as
\begin{subequations} 
\label{eq:gpsoc:1}
\begin{align}
\mathrm{i} \frac{\partial \psi_{\uparrow} }{\partial t} = & \left[ -\frac{1}{2}\nabla^2 +V_{2D}(x,y) +  \alpha \vert \psi_{\uparrow} \vert^2 + \beta\vert \psi_{\downarrow} \vert^2\right] \psi _{\uparrow} \notag \\  & 
- \Lambda_{+}^{SO} \psi_{\downarrow}, \label{eq:gpsoc:1-a} \\
\mathrm{i} \frac{\partial \psi_{\downarrow} }{\partial t} =& \left[ -\frac{1}{2}\nabla^2+V_{2D}(x,y) + \beta \vert \psi_{\uparrow} \vert^2+ \alpha \vert \psi_{\downarrow} \vert^2  \right] \psi _{\downarrow} 
\notag \\ & 
- \Lambda_{-}^{SO}  \psi_{\uparrow},  \label{eq:gpsoc:1-b}
\end{align}
\end{subequations}%
where $\nabla^2 = \partial_x^2 + \partial_y^2$, $V_{2D}(x,y) = \left( \lambda^2 x^2 + \kappa^2 y^2\right)/2$ is the harmonic trap potential and $\Lambda_{\pm}^{SO} = \left(k_L \left(\mathrm{i}\partial_x \pm\partial_y  \right) + \vert\Omega\vert \right)$. In the above equations (\ref{eq:gpsoc:1}), length is measured in units of harmonic oscillator length $a_{0}=\sqrt{\hbar/(m\omega_{\perp})}$, time in the units of $\omega^{-1}_{\perp}$, and energy in the units of $\hbar\omega_{\perp}$.  The  parameters $\alpha$ = $\sqrt{8\pi\eta} N a/ a_{0}$ and $\beta = \sqrt{8\pi\eta} N a_{\uparrow \downarrow} / a_{0}$ represent intra- and inter-species contact interaction strengths, respectively. The Rashba SO coupling and the Rabi coupling parameters have been rescaled as $k_L = k_L'/a_{0} \omega_{\perp}$ and  $\Omega = {\Omega'}/\omega_{\perp}$, respectively, while the wave function is rescaled as $ \psi_{\uparrow, \downarrow} = \Psi_{\uparrow, \downarrow} a_{0}^{3/2} /\sqrt{N}$.  We consider the Rabi coupling as $\Omega=\vert\Omega\vert e^{\mathrm{i}\theta}$ that minimizes the energy when $\Omega=-\vert\Omega\vert$ for $\theta=\pi$~\cite{Abad2013}. The wave functions are subjected to the following normalization condition,
\begin{align}
 \int_{-\infty}^{\infty}  \int_{-\infty}^{\infty}  \left( \vert \psi_\uparrow \vert^2 + \vert \psi_\downarrow \vert^2 \right) \, dx \, dy = 1,
\end{align}
The stationary state solution of the wave function is given by 
\begin{align} \label{chem}
\psi_{j}(x,y) = \big(\psi_{jR} + \mathrm{i} \psi_{jI} \big) e^{-\mathrm{i} \mu_{j} t}
\end{align}
 where, $j\in \{\uparrow, \downarrow\}$, $\psi_{jR}$ and $ \psi_{jI}$ are the real and imaginary part of the stationary wave function respectively,  $\mu_{\uparrow, \downarrow}$ are the chemical potential of spin-up and down components respectively. Now using Eqs.~(\ref{eq:gpsoc:1-a}), (\ref{eq:gpsoc:1-b}) and (\ref{chem}) the ground state energy of the Rashba SO coupled BECs can be obtained as
\begin{align}\label{eq:enum}
E_{num} = \sum_{j = \uparrow, \downarrow} \frac{\iint \left( E_{j}^{2C} + E_{j}^{SO} \right) \, dx \, dy}  {\iint  \psi_{jR}^2\, dx \, dy}
\end{align}
The detailed form of the $E_{j}^{2C}$ and  $E_{j}^{SO}$ are  given in the Appendix~{\ref{app-ene}}.

At first, we provide a detailed analysis on single-particle spectrum both in $k_x$ and $k_y$ momentum directions. Next, we present the analytical and numerical studies of the collective excitation spectrum to investigate the stability of the ground states obtained from the coupled GP equations~(\ref{eq:gpsoc:1-a}) and (\ref{eq:gpsoc:1-b}) as the small fluctuation is added in the ground state. It is followed by the detailed numerical analysis of the dynamics of the ground states. 

\section{Single-particle spectrum}
\label{sec:3}
In this section, first we present the calculation of two-component coupled GP equations with Rashba SO coupling for non-interacting BEC without any trapping potential, which gives the ``single particle spectrum''. Following this we include the intra- and inter-species nonlinear contact interactions with small perturbation, which gives rise to the ``excitation spectrum'' of the coupled system. This also includes the interesting features that appear due to interplay of Rashba SO ($k_L$) and Rabi couplings ($\Omega$).

Let's consider the Eqs.~(\ref{eq:gpsoc:1-a}) and (\ref{eq:gpsoc:1-b}) in the absence of  trap and contact interaction strengths, (i.e, $V= \alpha = \beta=0$) and use the plane wave solution $\psi_{\uparrow, \downarrow} = \phi_{\uparrow,\downarrow}\textrm{e}^{\mathrm{i}(k_x x + k_y y - \nu t)}$ in the computational basis, we get
\begin{align} \label{spdm}
\nu = 
\begin{pmatrix}
\frac{1}{2} \left(k^2_x+ k^2_y\right) & k_L \left(k_x - \mathrm{i} k_y \right)-\vert\Omega\vert \\
k_L \left(k_x + \mathrm{i} k_y\right) - \vert\Omega\vert  & \frac{1}{2} \left(k^2_x + k^2_y\right)  
\end{pmatrix},
\end{align}
which gives the single particle energy spectrum $\nu$ from equation (\ref{spdm}) as,
\begin{align}\label{spdr}
\nu(\mathbf{k})_{\pm} = \frac{1}{2} \left(k^2_x + k^2_y\right) \pm \left(\sqrt{\left(k_L k_x - \vert\Omega\vert \right)^2 + k_L^2 k_y^2} \right).
\end{align}

The single particle spectrum has two branches. First is the positive branch ($\nu_{+}$) which always have the single minimum and second is the negative branch ($\nu_{-}$) that makes transition from the single minimum to the double minima as the SO coupling strength ($k_L$) is increased for a fixed Rabi coupling. In what follows we focus our study on analyzing the negative branch as it exhibits transition between the single minimum to the double minima. We analyze the spectrum with respect to $k_x$ and $k_y$  momentum directions which will be useful to analyse the phase transition clearly.
\begin{figure}[ht] 
\centering\includegraphics[width=0.99\linewidth]{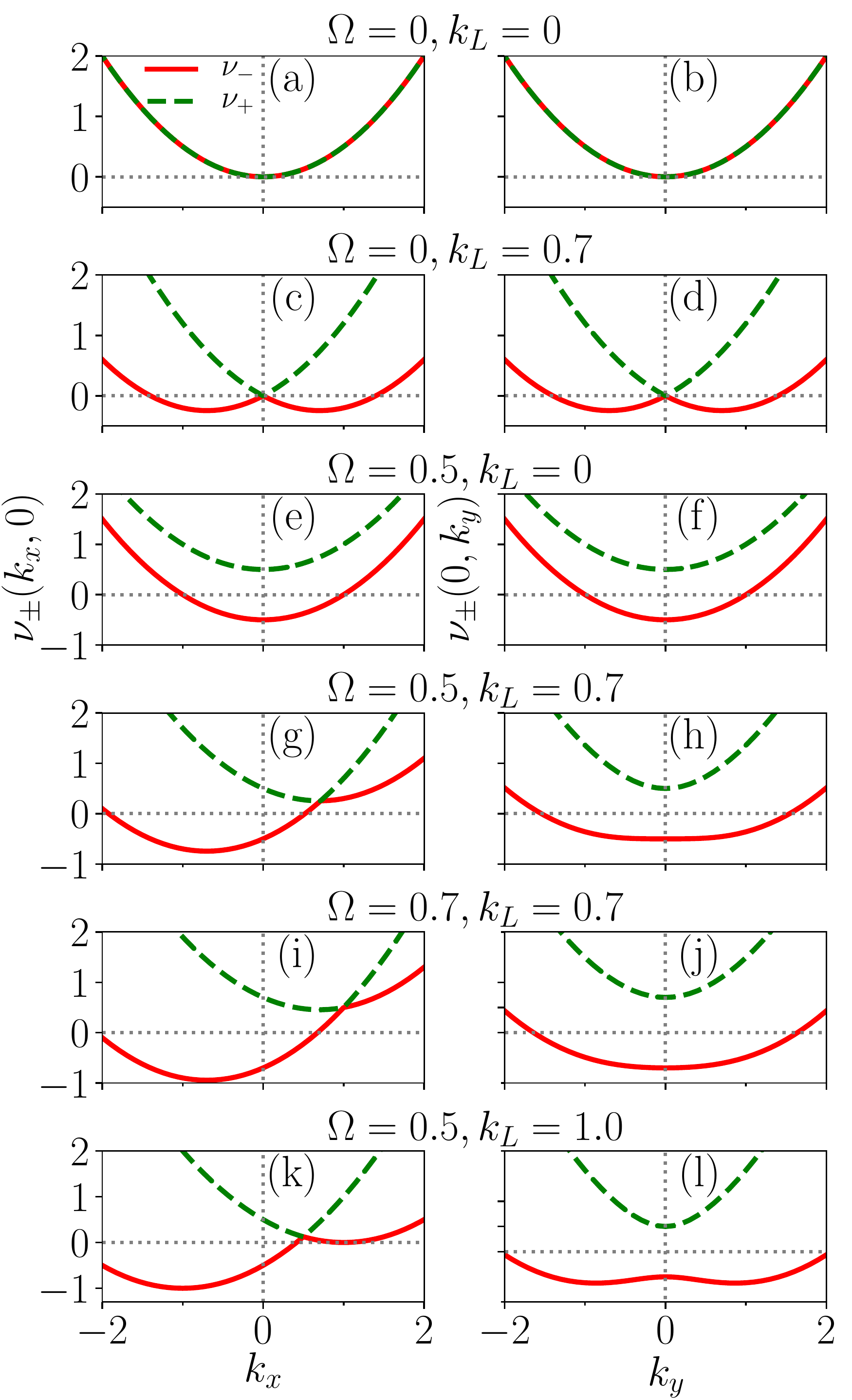}
\caption{Single particle energy spectrum in the $\{k_x,k_y\}$ momentum space for different set of coupling parameters indicated in the figure. Left column indicates $\nu_{\pm}(k_x,0)$ with $k_y =0$ and right column shows $\nu_{\pm}(0,k_y)$ for fixed $k_x =0$. Solid red line shows $\nu_{-}(k_x, k_y)$ and green dashed line is for $\nu_{+}(k_x, k_y)$. }
\label{fig1} 
\end{figure}

{\it {In $k_x$-direction:}} for zero Rabi and SO couplings ($\Omega=k_L=0$) the spectrum is a non-degenerate parabolic single-particle dispersion spectrum (Fig.~\ref{fig1}(a)) which can also be seen from the Eq.~\ref{spdr}. However, the spectrum exhibits double minima characteristics for finite values of the SO coupling strength ($k_L$). The minima are located at the  position $k_x = \pm k_L$ (Fig.~\ref{fig1}(c)). The value of energy minima increases with the SO coupling strength as $\pm k_L^2 /2$. For nonzero Rabi coupling ($\Omega\neq0$) and $k_L = 0$ the spectrum exhibits a single minimum and develops an energy gap between $\nu_{\pm}$ which is equal to $2\Omega$ (Fig.~\ref{fig1}(e)). However, for non-zero $k_L$ spectrum develops asymmetric double minima  as shown in Fig.~\ref{fig1}(g)).  As the Rabi coupling is increased, the system attains the minimized energy state. For example, with $k_L=0.7$, and $\Omega = 0.5$, the spectrum have a global minimum with $\nu_{-} = -0.745$ (See Fig.~\ref{fig1}(g)). Further increase in Rabi coupling from $\Omega = 0.5$ to $0.7$ results in lowering the energy state to $\nu_{-} = -0.945$ (in Fig.~\ref{fig1}(i)).  As $\Omega$ is increased further, the spectrum develops a global minimum and the corresponding energy varies as $k_L^2/2 + \Omega$. 

For the fixed Rabi coupling $\Omega = 0.5$, as we change the SO coupling strength from $k_L=0.7$ to $k_L=1.0$ we find that the  energies change from single minimum, i.e., $\nu_{-} = -0.745$ to double minima $\nu_{-} = \{-1.0,0\}$ respectively. It suggests that in general the atoms will get condensate in the lowest energy state, i.e, global minimum (See Fig.~\ref{fig1}(i)). While increasing $k_L$ we found two minima with opposite wavevectors which will form stripe like patterns (See Fig.~\ref{fig1}(k)).

{\it {In $k_y$-direction:}} The spectrum exhibits two minima in the absence of Rabi coupling with a finite $k_L$ (See Fig.~\ref{fig1}(d)). The spectrum typically known as {\it{Rashba ring}} in 2D momentum space~\cite{Wang2010,Jin2014}. As the Rabi coupling is increased for fixed $k_L$ a transition from the double-minima state to the single minimum state is observed at critical $\Omega (\approx k_L^2 )$ while symmetry of the system remains unchanged. 
We find that the energy gap between these two energy spectrum is $2\Omega$ as shown in the Figs. ~\ref{fig1}(h, j). The energy of the system gets lowered as Rabi coupling is increased. As for an example, as the coupling parameter is changed from $\Omega = 0.5$ to $\Omega =0.7$ for a  fixed $k_L=0.7$, the spectrum energies change from $\nu_{-} = -0.5$ and $\nu_{-} = -0.7$ (See Figs.~\ref{fig1}(h) and \ref{fig1}(j)). The energy strongly depends on Rabi coupling strengths for the cases $\Omega > k_L^2$. However, for $\Omega < k_L^2$ the energy depends on SO coupling strengths.

As we perform the energy comparison between $k_x$ and $k_y$ direction we find that in presence of the Rabi coupling the energy have the lowest value in the $k_x$ direction indicating the breaking of the rotational symmetry of the single-particle energy spectrum. This feature suggests that plane wave spectrum will get sustained with finite momentum for $\Omega > k_L^2$. However, for $\Omega < k_L^2$ stripe wave will exist. We do not observe any zero momentum (ZM) phase for $k_L\neq 0$ which was realized in one dimensional SO coupled BEC~\cite{Li2012, Ravisankar2020} and in two-dimensional  BECs~\cite{Jin2014, Bhuvaneswari2018} with nonlinear contact interactions. 

In the following section, we investigate the effect of couplings on the excitation spectrum of the SO coupled BECs.

\section{Analysis of the excitation spectrum of Rashba SO coupled BEC\lowercase{s}}
\label{sec:4}
In this section we present our analytical and numerical investigation of the excitation spectrum of the coupled BEC with SO and Rabi couplings. In 1941 Landau initiated the concept of elementary excitation to explain the  Superfluid behaviour in $^4$He. The mathematical derivation of excitation spectrum for the Bose gases was given by Bogoliubov in 1947~\cite{Bogolyubov1947}. Collective excitation spectrum of BECs gives the basic information about the dynamics of the quantum systems such as superfluid helium \cite{Zilsel1950}, superconductors \cite{Rickayzen1959}, phonon-like excitation observed in $^{87}$Rb atom in optical trap~\cite{Jin1996}, and magnetic trap~\cite{Mewes1996}, etc.  Here our aim is to investigate the excitation spectrum analytically which at the later part of the paper will be complemented with the numerical simulation.  At first, we analytically transform the pure ground state wavefunction ($\psi_{\uparrow, \downarrow}$) by adding a small perturbation ($\delta\psi_{\uparrow, \downarrow}$). After a direct algebraic manipulations, the system has four different excitation branches, namely, two positive and two negative branches. We analyze the characteristic of those energy branches by varying the SO and Rabi coupling parameters and further investigate the stability of the different modes. 
\subsection{Analytical and numerical description of excitation spectrum}
In order to understand the stability of our system, we calculate the excitation spectrum of plane wave solutions using Bogoliubov theory. Let's assume that the total density of the system  is $n = n_{\uparrow}+n_{\downarrow}$ and the chemical potential is $\mu$. Therefore the stationary state evolution can be written as \cite{Goldstein1997,Abad2013}:
\begin{align}
\Psi_j & = \mathrm{e}^{-\mathrm{i} \mu t} \left[ \psi_j +  \delta\psi_{j} \right], \label{eq:perturb} \\
\delta\psi_{j} & = u_j \mathrm{e}^{\mathrm{i} (k_x x + k_y y - t \omega )} + v^*_j \mathrm{e}^{-\mathrm{i} (k_x x + k_y y - t \omega^*)}, \label{eq:perturb-b}
\end{align}
where $\psi_j = \sqrt{n_{j}} \mathrm{e}^{\mathrm{i} \varphi_j}$, $j = (\uparrow, \downarrow)$ is the ground state wave functions, $u_j$ and $v_j$ are the amplitudes, $n_{j}$ and $\varphi_j$ are the density and phase respectively. The Bogoliubov coefficients $u$'s and $v$'s could be obtained by substituting Eq.~(\ref{eq:perturb}) in Eqs.~(\ref{eq:gpsoc:1-a}) and (\ref{eq:gpsoc:1-b}). Therefore, we have
\begin{align}
\label{eq:gpsoc:4} 
\omega 
\begin{pmatrix}
   u_{\uparrow} &
   v_{\uparrow} &
   u_{\downarrow} &
   v_{\downarrow}
\end{pmatrix}
^{T}
=
\mathcal{L} 
\begin{pmatrix}
   u_{\uparrow} &
   v_{\uparrow} &
   u_{\downarrow} &
   v_{\downarrow} 
\end{pmatrix}
^{T}
\end{align}
where the superscript $T$ denotes transpose of matrix and 
\begin{align}
\label{eq:gpsoc:5} 
\mathcal{L} = 
\begin{pmatrix}
 f(n_\uparrow, n_\downarrow) & \alpha n_{\uparrow} & L_{13} & \beta \sqrt{n_{\uparrow} n_{\downarrow}} \\
 -\alpha n_{\uparrow} & - f(n_\uparrow, n_\downarrow)& -\beta \sqrt{n_{\uparrow} n_{\downarrow}} & -L_{24} \\
L_{31} & \beta \sqrt{n_{\uparrow} n_{\downarrow}} &  g(n_\downarrow,n_\uparrow)& \alpha n_{\uparrow} \\
 -\beta \sqrt{n_{\uparrow} n_{\downarrow}} & -L_{42} & -\alpha n_{\downarrow}  & -g(n_\downarrow,n_\uparrow) 
\end{pmatrix},
\end{align}
and other coefficients are given in the Appendix~\ref{app-matrix}. The normalization condition yields
\begin{align}
\iint (\vert u_{j} \vert^2 - \vert v^*_{j}\vert^2 )\, dx \, dy = 1.
\end{align}

The simplified form of Bogoliubov-de-Gennes (BdG) equation under the condition $\text{det} \, \mathcal{L} =0$ will have the form for the interacting case with $n_{\uparrow} = n_{\downarrow} = 1/2$ as 
\begin{align}
\label{spectrum-a}
\omega^4 + b \omega^2 + c \omega +d  = 0
\end{align}%
By direct mathematical manipulation of the equation~(\ref{spectrum-a}), we obtained four dispersion relations. The complicated expressions of the coefficients ($b, c, d$) are given in the Appendix~\ref{app-coeff}.

Following this we corroborate the analytical results for the excitation spectrum by numerically solving the BdG equations from which we also obtain the eigenvectors as a function of $k_x$ and $k_y$. First we consider a $[-1000:1000][-1000:1000]$ grid in real space with step size $h_x=h_y=0.05$. Then we use the Fourier collocation method where we numerically perform the Fourier transformation of BdG equations and obtained a truncated reduced BdG matrix, which was subsequently diagonalized using the LAPACK package~\cite{Anderson1999}. In momentum space we consider $[-50:50][-50:50]$ modes in $k_x$, $k_y$ directions with a grid step size of $h_{k_x}=h_{k_y}=0.0628$.

\subsection{Comparison of excitation spectrum with and without interactions}
\begin{figure}
\centering\includegraphics[width=0.99\linewidth]{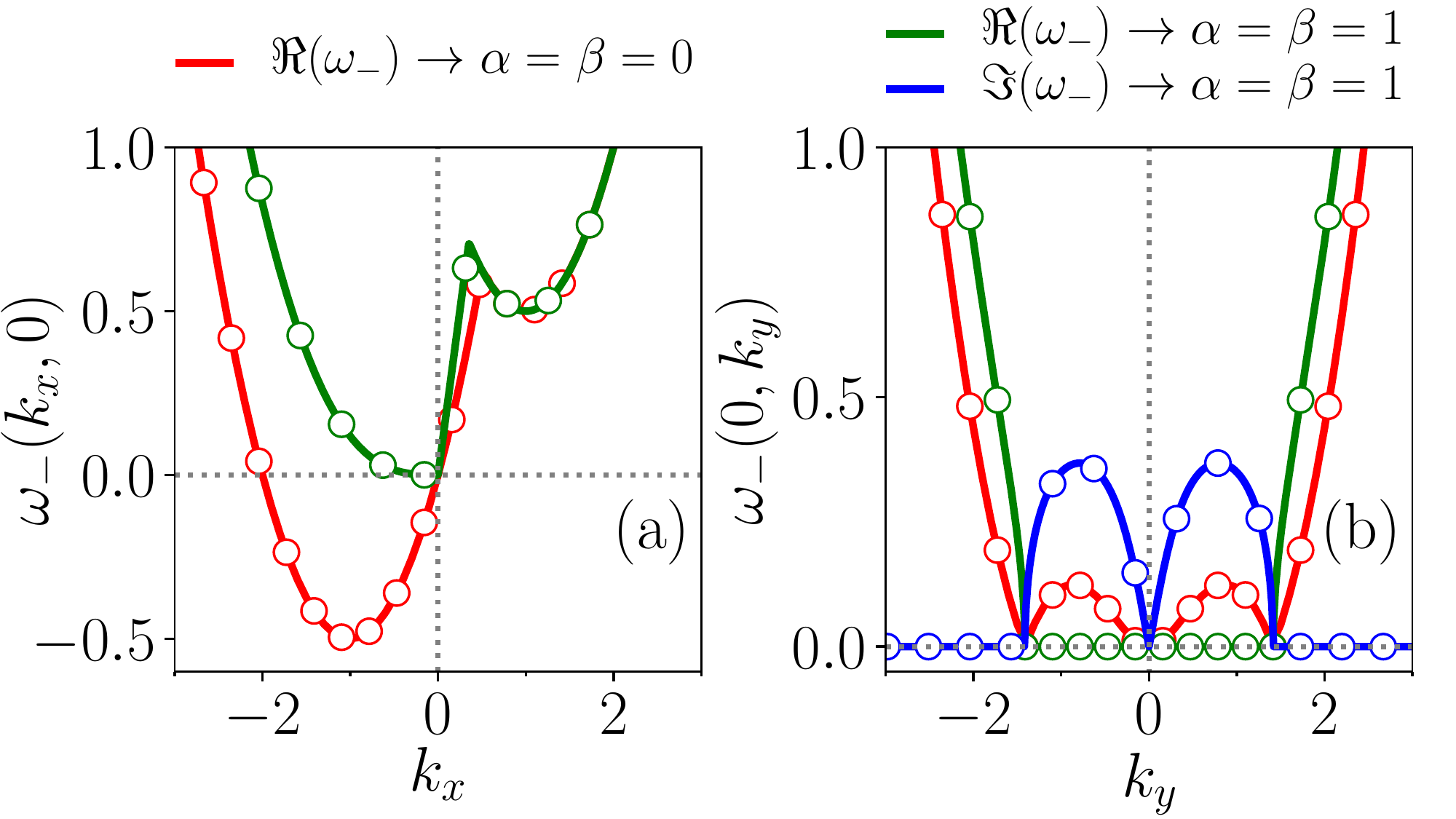}
\caption{Collective excited dispersion of $\Re(\omega_{-})$ (solid red line) for non-interacting  and of $\Re(\omega_{-})$ (solid green line) and $\Im(\omega_{-})$ (solid blue line) for interacting ($\alpha=\beta=1$) case. The parameters are $\Omega=0.5$, $k_L=1$. (a) Variation of negative spectrum along the $k_x$ direction for $k_y=0$ and  (b) Variation of negative spectrum along the $k_y$ direction for $k_x=0$. Solid lines represent the analytical results obtained from BdG Eq.~(\ref{spectrum-a}) and open circles denote the numerical solution of Eq.~(\ref{eq:gpsoc:4}).}
\label{fig2}
\end{figure}

In the following we analyze the effect of the couplings on the stability of the negative branch of the energy spectrum. Note that the negative 
eigenenergy of the excitation spectrum implies that the system is energetically unstable, while, the imaginary or complex eigenenergies indicate the dynamical instability~\cite{Ozawa2013}. Bogoliubov-de-Gennes excitation spectrum obtained from equation~(\ref{spectrum-a}), without SO coupling ($k_L=0$) is similar as obtained in Ref.~\cite{Abad2013, Goldstein1997}. Apart from this the dispersion relation has two dimensional Rashba SO and Rabi couplings. Fig.~\ref{fig2} shows system of collective BdG excitation  spectrum. The comparison between the noninteracting (solid red lines) and  interacting ($\alpha = \beta = 1$) cases (shown with the green and blue lines) indicate that the former case does not have any imaginary part while later case have the imaginary part. This particular features suggests that the metastability of the plane wave phase of SO coupled BECs is destroyed in the $k_y$-direction which makes the system dynamically unstable indicating the system lacks any superfluid behaviour~\cite{Zhu2012}. 

\begin{figure}
\centering\includegraphics[width=0.99\linewidth]{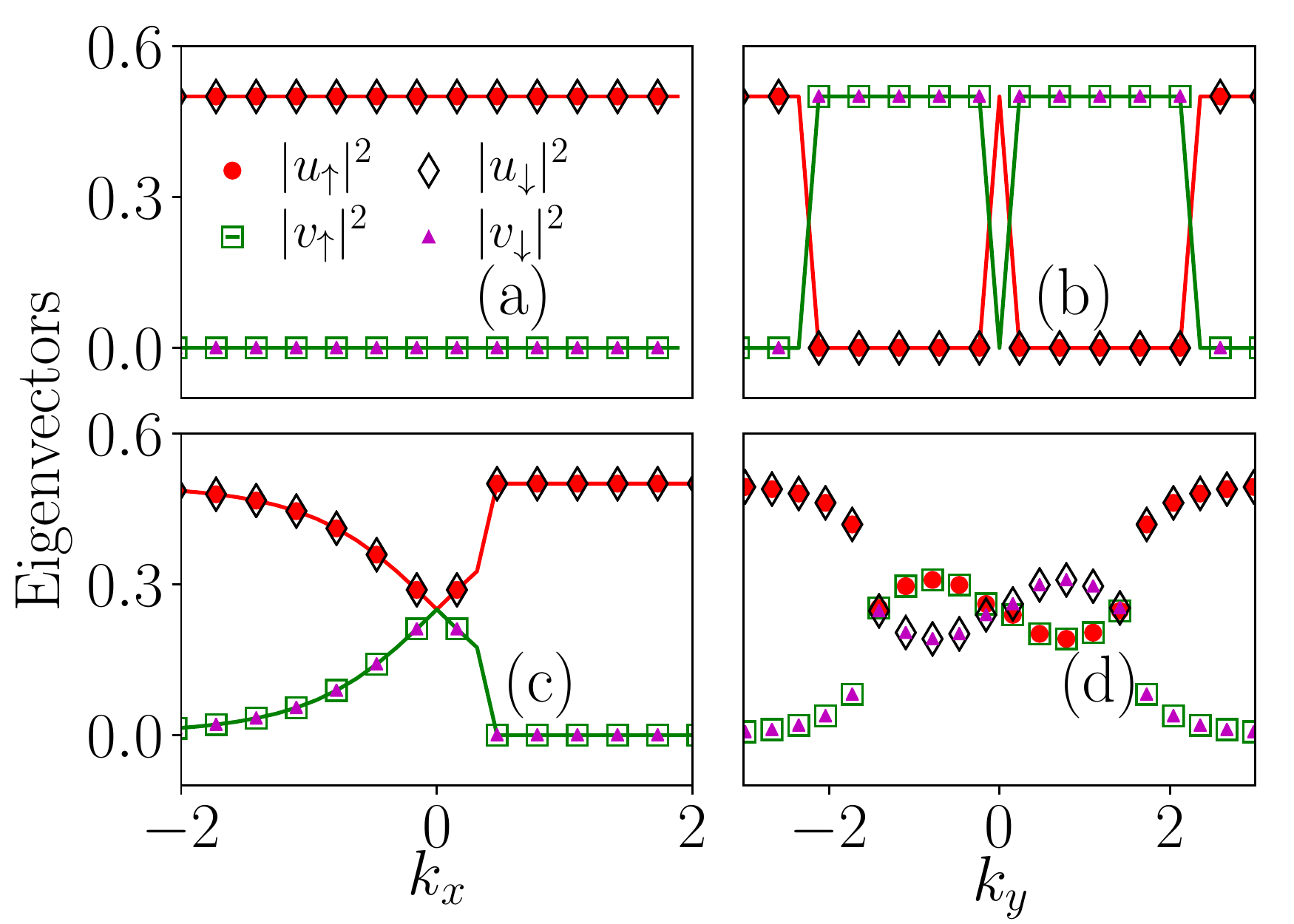}
\caption{The eigenvectors obtained by solving Eq.~(\ref{eq:gpsoc:4}): Red dots for $\vert u_{\uparrow}\vert$ component, Black open diamond for $\vert u_{\downarrow}\vert$, green open square for $\vert v_{\uparrow}\vert$ and magenta triangle for $\vert v_{\downarrow}\vert$.   Top row is for the noninteracting case and bottom row is for the interacting case. Left column shows the variation of eigenvectors along the $k_x$ directions and right column those variation along the $k_y$ directions. All the other parameters are same as in Fig.~\ref{fig2}. For non-interacting case plane wave, density mode dominates for all range of the wave number, while for the interacting case a transition from the density to spin mode occurs where the negative spectrum becomes complex in $k_y$ momentum direction. }
\label{fig2eigvec}
\end{figure}%

In Fig.~\ref{fig2eigvec}, we plot the eigenvectors corresponding to the  eigenenergy spectrum as shown in the Fig.~\ref{fig2}. For non-interacting case the in-phase of the eigenvector components for all wave number indicate the presence of  only density like modes (in-phase) in both momentum directions (See Fig.~\ref{fig2eigvec}(a,b)). However, for interacting case   ($\alpha = \beta = 1$) presence of some complex patterns are observed. There is transition from the density like modes (in-phase) to the spin-like mode (out-phase) happens in $k_y$ direction as shown in the Fig.~\ref{fig2eigvec}(c,d). We find that both eigen components of the eigenvectors $u$'s and $v$'s are in-phase in the $k_x$ direction. For $k_x \approx 0$ eigenvectors  approach towards each other and have equal values at $k_x = 0$, which indicates the presence of phonon mode. Beyond this there is sudden increase in the value of  the eigenvectors. At finite $k_x$ again there is a change in the curve that corresponds to the maxon point which is followed by returning to the  density-like mode. 
In $k_y$ direction, we found two types of behaviour. First up to the $\Re(\omega_{-})$ the $u$'s and $v$'s are in-phase, while the presence of $\Im(\omega_{-})$ changes $u$'s and $v$'s in out-phase which indicate the presence of spin-like mode~\cite{Abad2013}.  
\subsection{Effect of Rabi coupling on the excitation spectrum}
Theoretically it was noticed that excitation spectrum of quasi-one-dimensional Raman SO coupled BECs consists of roton like minimum for the case of finite Rabi coupling \cite{Martone2012,Zheng2013} which was also confirmed experimentally in~\cite{Khamehchi2014,Ji2015}. In this section we present the effect of Rabi coupling on the excitation spectrum in two dimension coupled BECs.
\begin{figure}[!ht]
\centering\includegraphics[width=0.99\linewidth]{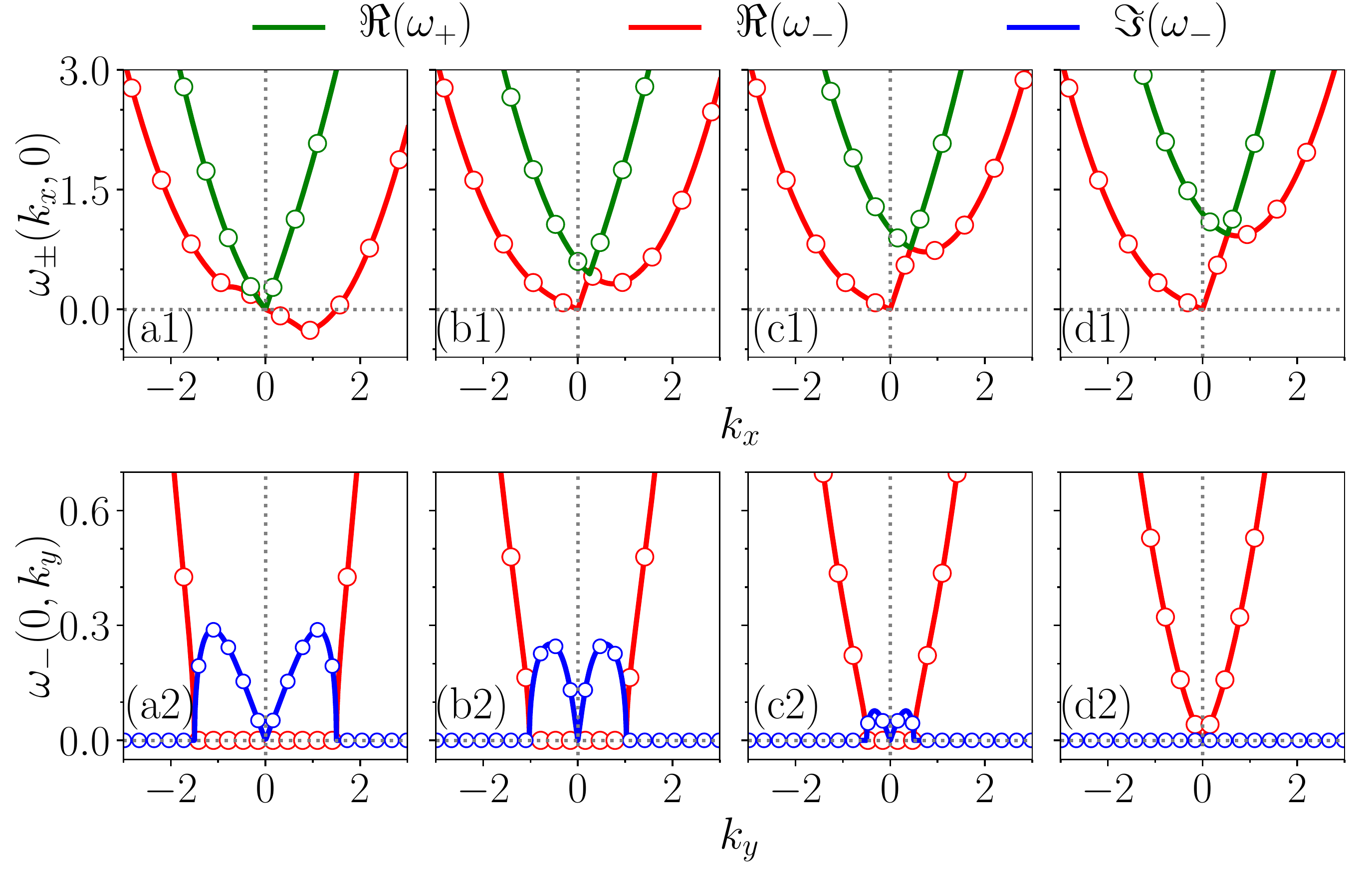} 
\caption{Excitation spectrum in $k_x$ (first row) and $k_y$ (second row) momentum directions for different $\Omega$. $\Omega$ varies in column (a)-(d) = ($0, 0.3, 0.5, 0.6$) and other fixed parameters are $k_L = 0.75$, $\alpha = \beta = 1$. First row: Solid red line and solid green line indicate $\omega_{-}(k_x, 0)$ and $\omega_{+}(k_x, 0)$ respectively; Second row:  Red solid line and solid blue line depicts that $\Re(\omega_{-}(0, k_y))$ and $\vert\Im(\omega_{-}(0, k_y))\vert$ respectively. Solid lines are analytical results from BdG Eq.~(\ref{spectrum-a}) and open circles are obtained numerically solving Eq.~(\ref{eq:gpsoc:4}). First row shows roton softening upon increase in $\Omega$, which reflects as a disappearance of the imaginary modes for high $\Omega$. }
\label{fig3}
\end{figure}
As Rabi coupling is decreased we find that beyond the critical value of $\Omega$ roton-like minimum starts softening, it develops a negative frequency, which indicates the appearance of instability in the system. We fix the SO coupling parameter to $k_L =0.75$ with $\alpha = \beta=1$, $k_y=0$, and vary the Rabi strength $\Omega$. We consider $\Omega=0,0.3,0.5,$ and $0.6$. In Fig.~\ref{fig3} we show the dispersion behaviour for different $\Omega$ by keeping the SO coupling parameter fixed. We notice the presence of minimum attributed to the  phonon, maxon and roton. At zero Rabi coupling ($\Omega = 0$) strength the excitation spectrum possesses negative frequency in $k_x$ direction indicating that the system is energetically unstable. However, in $k_y$ direction the system exhibits dynamical instability due to the presence of the complex excitation frequency. Owing to this feature at zero Rabi coupling system does not exhibit any superfluid behaviour~\cite{Zhu2012}. As the Rabi coupling strength is increased to $\Omega = 0.3$ the eigenfrequency suggests the presence of minimum related to phonon-maxon-roton in $k_x$ direction while in the transverse direction ($k_y$) it suggests the presence of the complex eigen-frequencies (See Fig.~\ref{fig3}(b1) and Fig.~\ref{fig3}(b2)). This behaviour indicates the dynamically unstable state for this SO coupling strength. For $\Omega = 0.5$, the system remains dynamically unstable, however, decrease in the roton minimum and amplitude of the imaginary frequency is observed (See Fig.~\ref{fig3}(c1) and Fig.~\ref{fig3}(c2)). 

As the Rabi coupling strength is increased beyond a threshold value $\Omega \geq 0.56$ the system does not show any complex frequency in $k_y$ direction indicating the stable behaviour. Only the presence of real frequency in the  $k_y$ direction indicates metastable state, also the axial symmetry is preserved (See Fig.~\ref{fig3}(d1) and Fig.~\ref{fig3}(d2)). Overall we find that the increase in the Rabi coupling strength for a fixed SO coupling leads the stabilization of the system.

\begin{figure}[!ht]
\centering\includegraphics[width=0.99\linewidth]{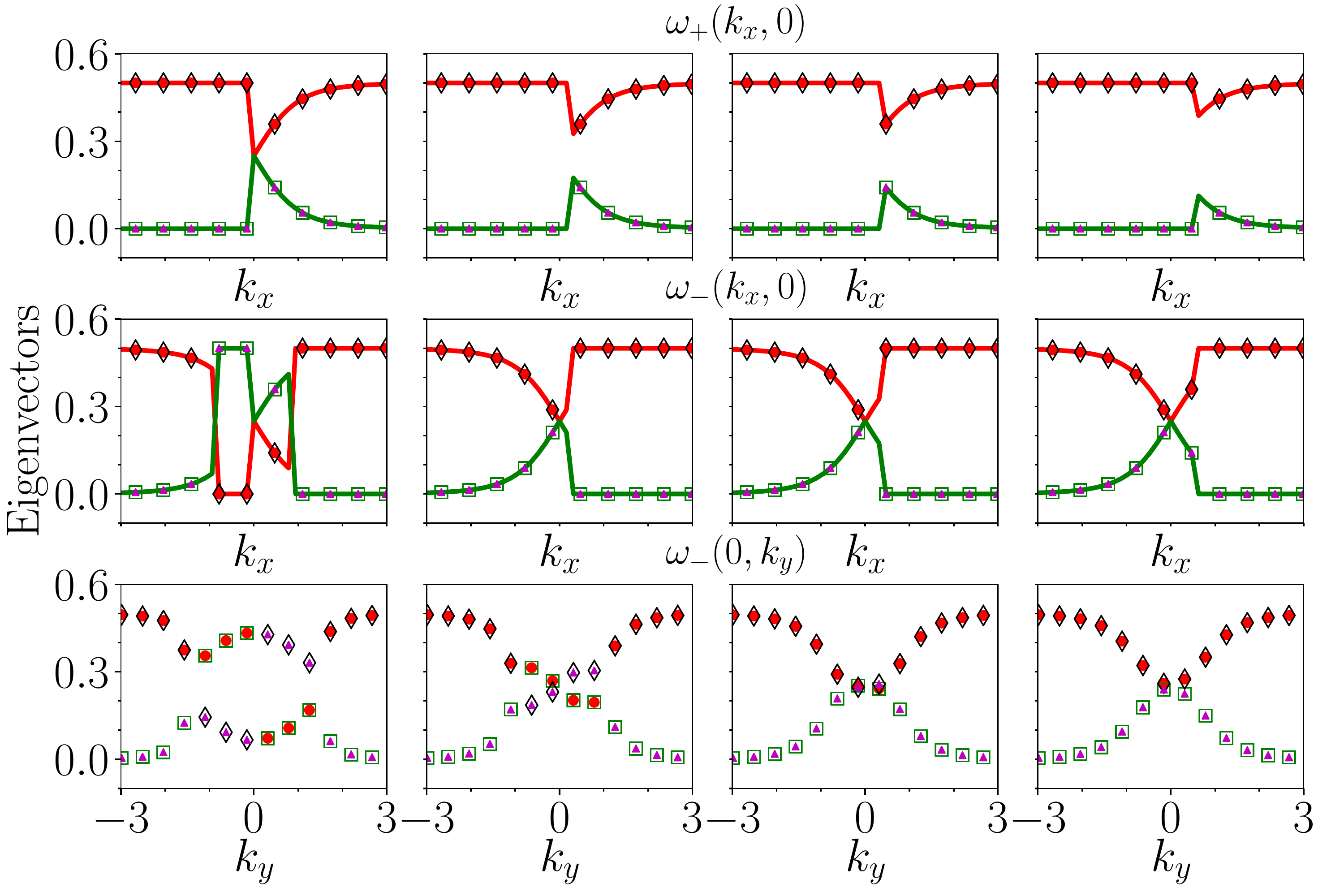}
\caption{The eigenvectors corresponding to the eigenspectrum of Fig.\ref{fig3}. The representation of the eigenvector components is same as in Fig~\ref{fig2eigvec}. Top panel represents the eigenvectors corresponding to $\omega_{+}(k_x,0)$ (solid green line), middle panel for $\omega_{-}(k_x,0)$ (solid red line) and bottom panel for $\omega_{-}(0, k_y)$. Maxon mode in the top panel, phonon-maxon mode in the middle panel and spin to density mode transition in the bottom panel are noticeable.}
\label{fig3egvec}
\end{figure}%
From the excitation spectrum we analyzed the phonon-maxon-roton modes. Now in order to get more insight about the stability of these modes we move to the analysis of the characteristics of the eigenvectors in the momentum space. Figs.~\ref{fig3egvec} shows the eigenvectors corresponding the eigen spectrum given in the Fig.~\ref{fig3}. The first row displays the eigenvectors corresponding to $\omega_{+}(k_x,0)$ while second and third row demonstrate the the eigenvectors  of $\omega_{-}(k_x,0)$, $\omega_{-}(0, k_y)$, respectively. For $\Omega = 0$ with $k_L = 0.75$ both eigenvectors in $k_x$ momentum direction are in-phase except near $k_x =0$. At $k_x=0$ both $u$'s and $v$'s becomes equal indicating the presence of phonon mode, and absence of maxon-mode.  However, for $\omega_{-}$ we observe a complicated spin-flipping behaviour in both $k_x$ and $k_y$ directions of eigenvectors, which is with respect to the excitation spectrum's negative and imaginary eigen frequency.  

Similar to the earlier studies~\cite{Goldstein1997,Abad2013, Tommasini2003, Recati2019} we observe the two branches in the spectrum. First one represents to the gapless density mode (in-phase mode) corresponding to the Goldstone mode with U(1) symmetry and second one denotes the gapped spin mode (out-of-phase mode). As we carefully analyze their corresponding eigenmodes we notice some important differences. At one hand for the density mode we obtain $|u_1|^2-|u_2|^2=|v_1|^2-|v_2|^2=0$, which in terms of spin language is an unpolarized mode, while at the other hand for the spin mode we find breaking of the $\mathbb{Z}_2$ symmetry, meaning that $|u_1|^2-|u_2|^2\neq0$ and $|v_1|^2-|v_2|^2\neq0$ that implies a polarized mode in the spin language~\cite{Recati2019}.

In Fig.~\ref{fig3}, we display the eigenmodes corresponding to the excitation spectra for different values of $\Omega$. For $\Omega<0.6$ (Figs.~\ref{fig3}(a)-\ref{fig3}(c)) the excitation spectra exhibit complex eigenvalues for small values of $k_y$ which indicates that the system is dynamically unstable for such parameter values. As we look at the eigenmodes of such excitation spectra we find that those instabilities can be attributed to the generation of the spin modes. Increase in value of $\Omega$ to $\Omega=0.6$ results complete suppression of such unstable spin mode and emergence of the density mode in the system. The same scenario appear to be present for the parameters of Fig.~\ref{fig5}, where increase of $k_L$ generates dynamical instabilities in the system and that manifest as the emergence of the spin modes.

\begin{figure}[!htp]
\centering\includegraphics[width=0.99\linewidth]{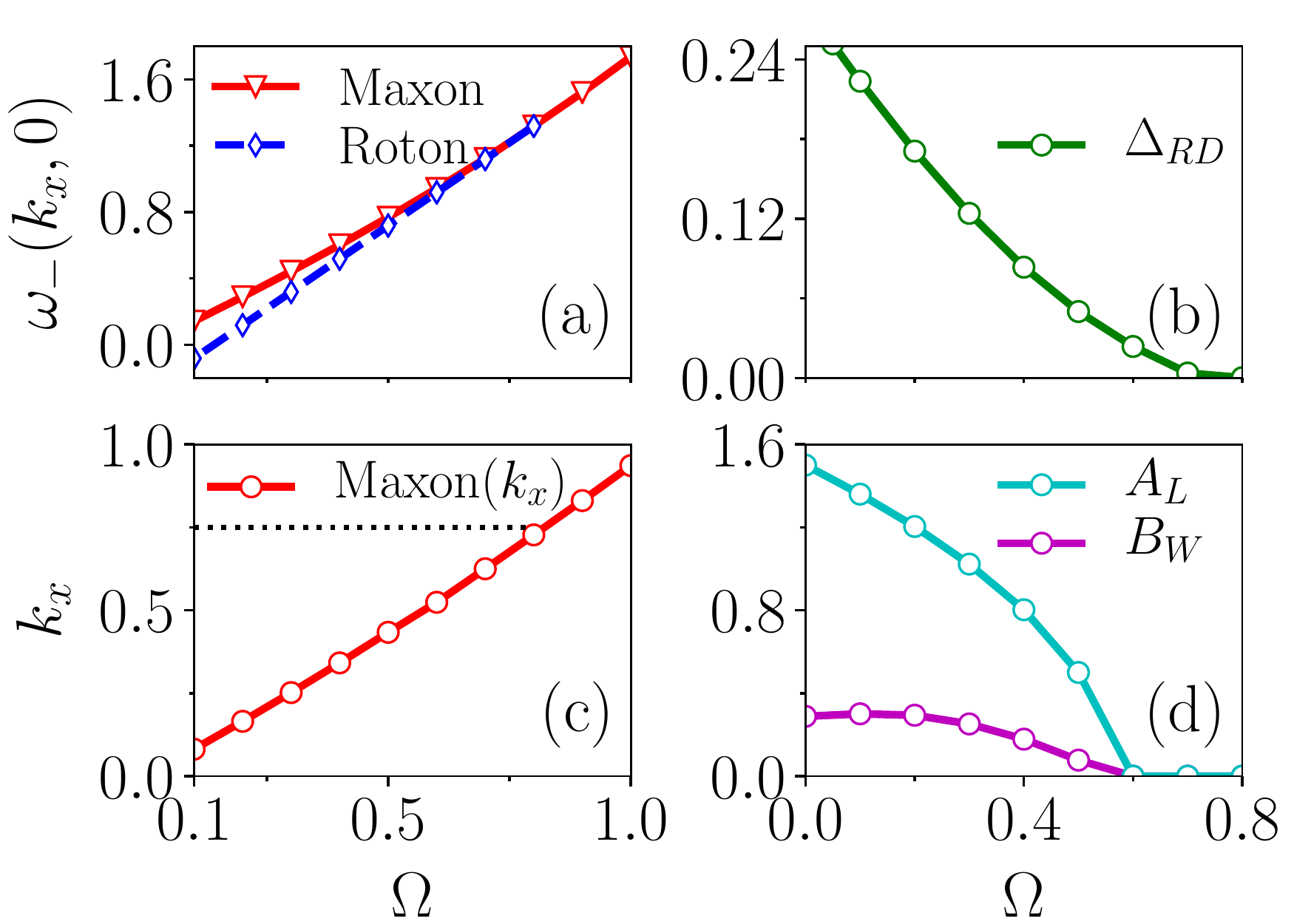}
\caption{Variation of different quantities with $\Omega$  for $\alpha = 1$, $\beta = 1$ and $k_L = 0.75$. (a) Variation of Maxon (red open inverted triangles) and Roton minima (blue open diamond), (b) Decay in roton depth ($\Delta_{RD}$), (c) Position of maxon (red open circles) and roton minima (black dotted lines) in $k_x$ momentum direction, (d) Amplitude loss in instability (cyan open circles) and corresponding bandwidth (magenta circles) in the $k_y$ momentum space. }
\label{fig4}
\end{figure}

We also analytically compute the effect of $\Omega$ with fixed parameters $k_L = 0.75, \alpha = \beta = 1$ on the maxon and roton minimum from the collective excitation spectrum. In Fig.~\ref{fig4}(a) solid red line with triangle, we show the maxon and blue line with diamond we indicate the roton minimum. The points of maxon and roton are well fitted with the straight line beyond the critical value of $\Omega$ both maxon-roton points are merged as shown in the Fig.~\ref{fig4}(a). As we understand that the difference between roton minimum and maxon gives the roton gap/depth ($\Delta_{RD}$) as displayed in Fig.~\ref{fig4}(b). As the Rabi coupling is increased, the roton minima disappears. Beyond the critical value only maxon in $k_x$ direction is observed. We find that the outcome of increase in Rabi coupling strength is the decay of roton gap/depth. 

In Fig.~\ref{fig4}(c) we show the positions of maxon and roton minimum positions along $k_x$ momentum space. We find that the roton minima are always present at $k_x = 0.75$, which appears to be similar as we  we observed for fixed Rashba SO coupling strength $k_L = 0.75$. However, upon looking the behaviour carefully we find that the position of maxon increases upon increase of the Rabi coupling strength and finally at $\Omega \approx 0.8$ the maxon and the roton gets merged. As a consequence of this above this Rabi coupling strength only maxon remains present in the system. In the similar way, we notice the loss in the amplitude of instability upon increasing of $\Omega$ denoted as $A_{L}$ (amplitude loss). In Fig.~\ref{fig4}(d) we plot the variation of  $A_{L}$ and instability bandwidth ($B_{W}$) with the Rabi coupling which clearly complement our results attributed to stabilization of the system beyond $\Omega \approx 0.8$.  

\begin{figure}[!ht]
\centering\includegraphics[width=0.99\linewidth]{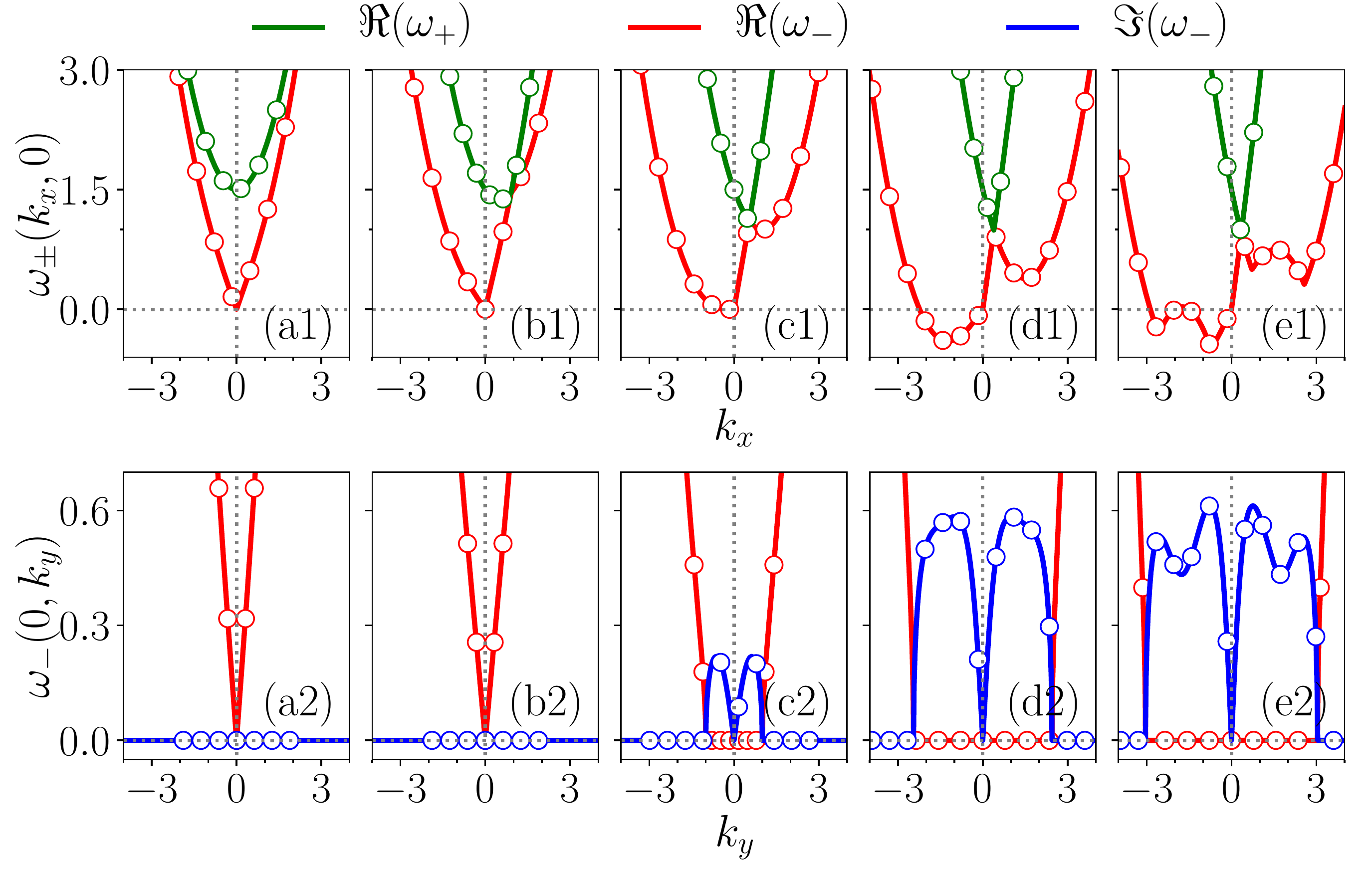} 
\caption{Variation of excitation spectrum in $k_x$ (first row) and $k_y$ (second row) momentum directions for different SO coupling strengths $k_L$, varies in column (a-e) = $(0, 0.5, 1.0, 1.5, 1.75)$. Other parameters are $\Omega =0.75, \alpha= \beta=1$. First row: Green and solid red lines indicate $\omega_{+}(k_x,0)$ and $\omega_{-}(k_x,0)$ respectively; Second row: solid red and blue lines show $\Re(\omega_{-}(0,k_y))$ and $\vert\Im(\omega_{-}(0,k_y))\vert$. Solid lines are analytical results from Eq.~(\ref{spectrum-a}) and open circles are numerically obtained from Eq.~(\ref{eq:gpsoc:4}). The increase in $k_L$ leads to generation of imaginary energy frequency modes in the $k_y$ direction.}

\label{fig5} 
\end{figure}
\subsection{Role of Rashba spin-orbit coupling ($k_L$)}
In this section, we investigate the effect of SO coupling strength on the excitation spectrum by fixing the other parameters, like, $\Omega=0.75$ and $\alpha = \beta = 1$. In absence of SO coupling ($k_L =0$) we find that the spectrum only have the real frequencies which are symmetric in both $k_x$ and $k_y$ directions, also exploring the zero momentum phase (see Fig.~\ref{fig5}(a1) and Fig.~\ref{fig5}(a2)). As the SO coupling is raised to $k_L = 0.1$ the maxon mode appears in the system. On further increase in $k_L$ leads to the generation of roton minimum in the spectrum as shown in Fig.~\ref{fig5}(b1). For $k_L > 0.86$, we find transition from stable to the unstable state which is quite evident by the presence of complex frequencies in $k_y$ directions  of $\omega_{-}(0, k_y)$ spectrum as shown in the Fig.~\ref{fig5}(c2). Further  increase in $k_L$ leads to increase in the magnitude of the instability (see Fig.~\ref{fig5}(d2)). For $k_L > 1.5$, we notice the negative frequency with two minima in $k_x$ direction,  which gives two stability bands in both side of $k_y$ direction (See Fig.~\ref{fig5}(e1) and Fig.~\ref{fig5}(e2)). We find that the role of SO coupling here is to gradually increase the phonon-maxon and roton minimum up to the critical value of $k_L$ in $k_x$ momentum space. Increasing $k_L$ beyond a critical value leads to the loss of phonon-maxon and roton minimum apart of having the negative energy that shows the system becomes energetically unstable and thus lacks any superfludity~\cite{Zhu2012}. 
\begin{figure}
\centering\includegraphics[width=0.99\linewidth]{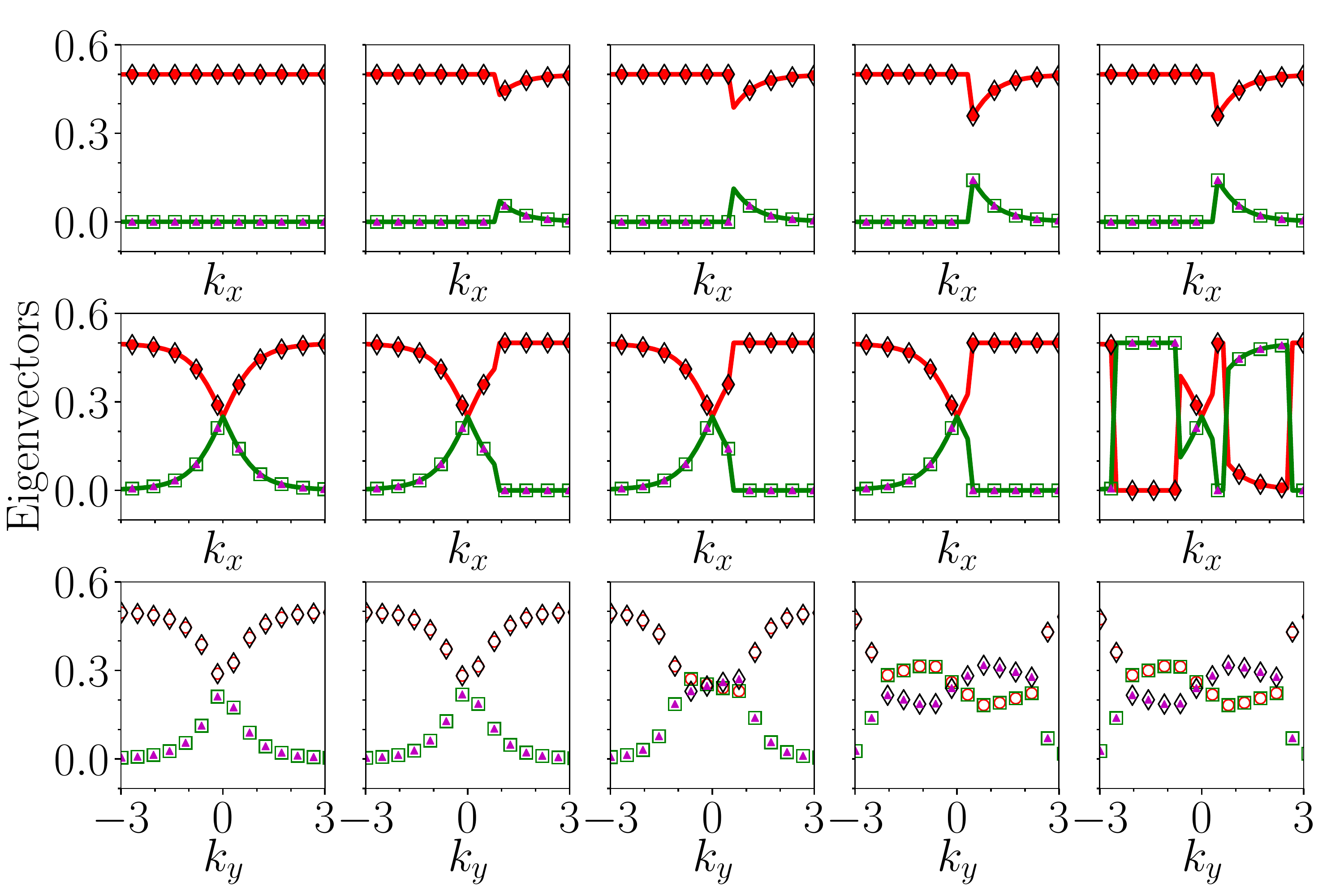}
\caption{Eigenvectors associated with Fig.~\ref{fig5}, which are obtained numerically from Eq.~(\ref{spectrum-a}). The representation of the eigenvector components is same as in Fig~\ref{fig2eigvec}. First row: $\omega_{+}(k_x,0)$; Second row: $\omega_{-}(k_x,0)$; Third row: $\omega_{-}(0, k_y)$. First row shows the evolution of maxon modes, second row displays phonon-maxon to complicated spin-flipping modes, third row clearly shows the transition from density to spin-mode. }
\label{fig5egvec}
\end{figure}%

In Fig.~\ref{fig5egvec} we plot the eigenvectors corresponding to eigen spectrum as shown in the Fig.~\ref{fig5}.  For $k_L \neq 0$ cases, in $\omega_{+}(k_x,0)$ spectrum's $u$'s ($v$'s) exhibits decreasing (increasing) trend to reach the point of maxon-mode which also shows the coupling point between local minimum of $\omega_{+}(k_x,0)$ and maxon of $\omega_{-}(k_x,0)$. In the $k_y$ momentum direction we initially notice the presence of density modes that get transmuted to spin-density modes as  $k_L$ is increased for fixed  $\Omega$ (Fig.~\ref{fig5egvec} third row). Through the eigenvector analysis we establish the existence of phonon-maxon modes. 
\begin{figure}[!htp]  
\centering\includegraphics[width=0.99\linewidth]{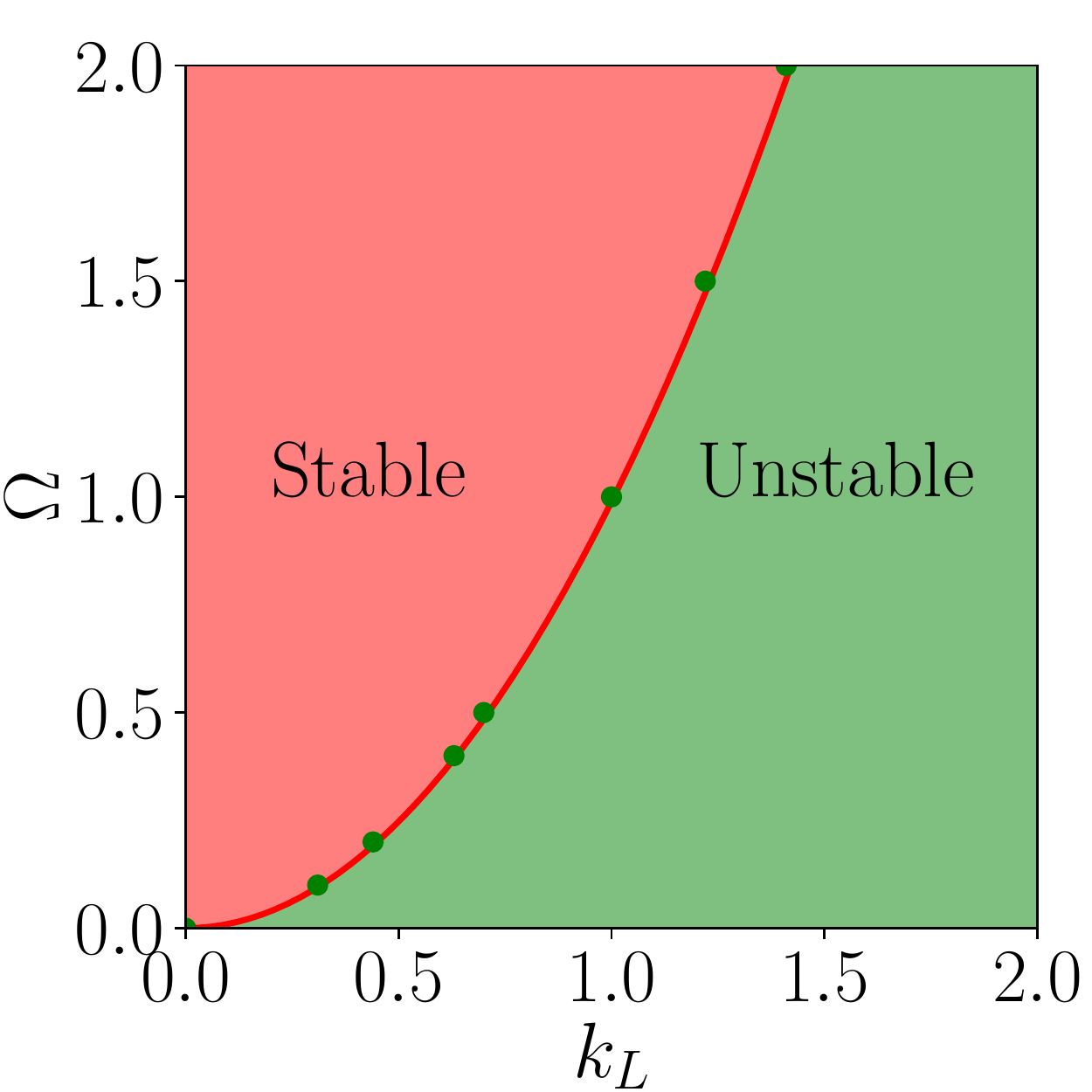}
\caption{Stability phase diagram illustrating the stable and unstable phases in $k_L-\Omega$ plane with fixed weak repulsive nonlinear contact interaction strengths, $\alpha = \beta = 1$. Green dots represent the phase transition points of unstable and stable boundary obtained from eigen spectrum using Eq.~(\ref{spectrum-a}).}
\label{fig6}
\end{figure}
\subsection{Stability phase diagram}
In order to provide a detailed picture of the stability of ground state phases we identify a phase transition from stable to unstable phase in $k_L-\Omega$ plane using excitation spectrum and illustrate this in the phase-diagram (see Fig.~\ref{fig6}). By simultaneously varying $\Omega$ and $k_L$ real and complex frequencies where present in the $k_y$ direction those critical points are noted for a fixed $k_x=0$. We show the critical points in the $k_L-\Omega$ plane with the green dots. We characterize the phase as stable one if the frequencies are real otherwise phase is denoted as unstable for the  complex frequencies~\cite{Goldstein1997, Ozawa2013, Zhu2012}. 

After having the fair understanding about the different phases of excitation spectrum of the coupled BEC using BdG equations now in the following section we complement our observation by directly solving the set of SO coupled GPE.
\section{Numerical simulation results}
\label{sec:5}
In this section, we present the numerical results which is obtained by solving the dynamical equations (Eqs.~\ref{eq:gpsoc:1}) of the coupled BEC. We use the imaginary-time propagation (ITP) method to obtain the ground state of the system. Subsequently, we evolve the ground state wave-function by using the conventional real-time propagation (RTP), where in both the methods, we adopted the split-step Crank-Nicholson scheme~\cite{Muruganandam2009, Young2016, Ravisankar2021, Muruganandam2021}. We have considered the grid sizes $400 \times 400$ with space steps $dx = dy = 0.1$, and time step $dt = 0.005$ is used.
Initially, we obtained the ground state wave-functions using ITP method with respect to stable ($k_L=0, \Omega=0.75$) and unstable ($k_L=1.5, \Omega=0.75$) regime of the stability phase diagram of Fig.~\ref{fig6}. Once the ground state wave-function is obtained, we then evolve it employing the RTP method, in which, we quench the system by reducing the trap strength as $\lambda = \kappa = 1.0 \to 0.5$ at $t = 0$.

\begin{figure}[ht]  
\centering\includegraphics[width=0.99\linewidth]{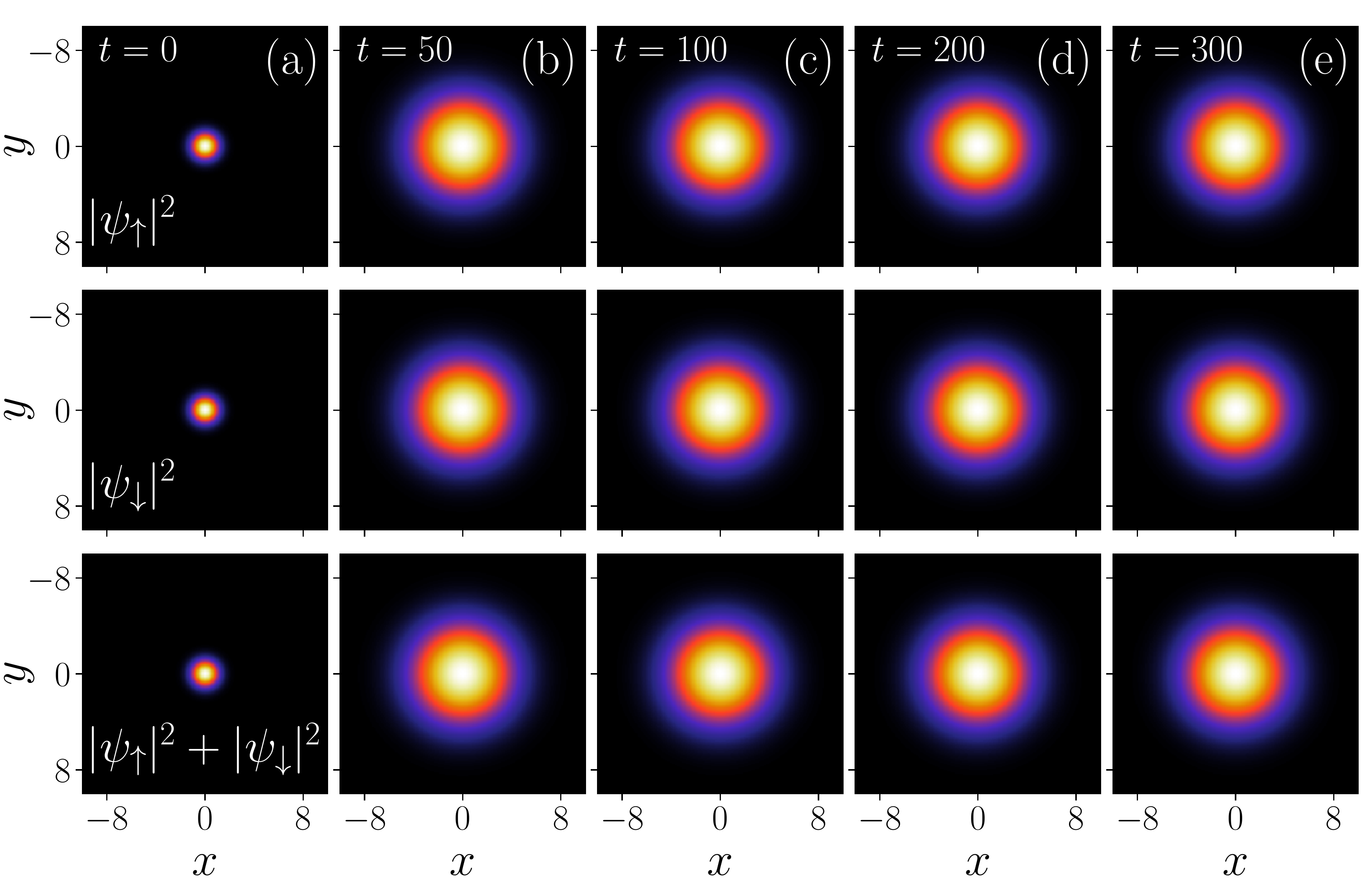}
\caption{Time evolution of the ground state density for $\Omega = 0.75$, $k_L = 0$ and $\alpha = \beta = 1$. First column (a) is the actual ground state wavefunctions obtained from imaginary-time propagation method at $t =0$. (b-e) are their dynamical density patterns at different time instant $t = 50, 100, 200, 300$ (column wise). Top row indicates spin-up ($\vert\psi_{\uparrow}\vert^2$) density, spin-down ($\vert\psi_{\downarrow}\vert^2$) in the middle, and total density ($\vert\psi_{\uparrow}\vert^2$ + $\vert\psi_{\downarrow}\vert^2$) in the bottom row.}
\label{fig7}
\end{figure}
\begin{figure}[ht]  
\centering\includegraphics[width=0.99\linewidth]{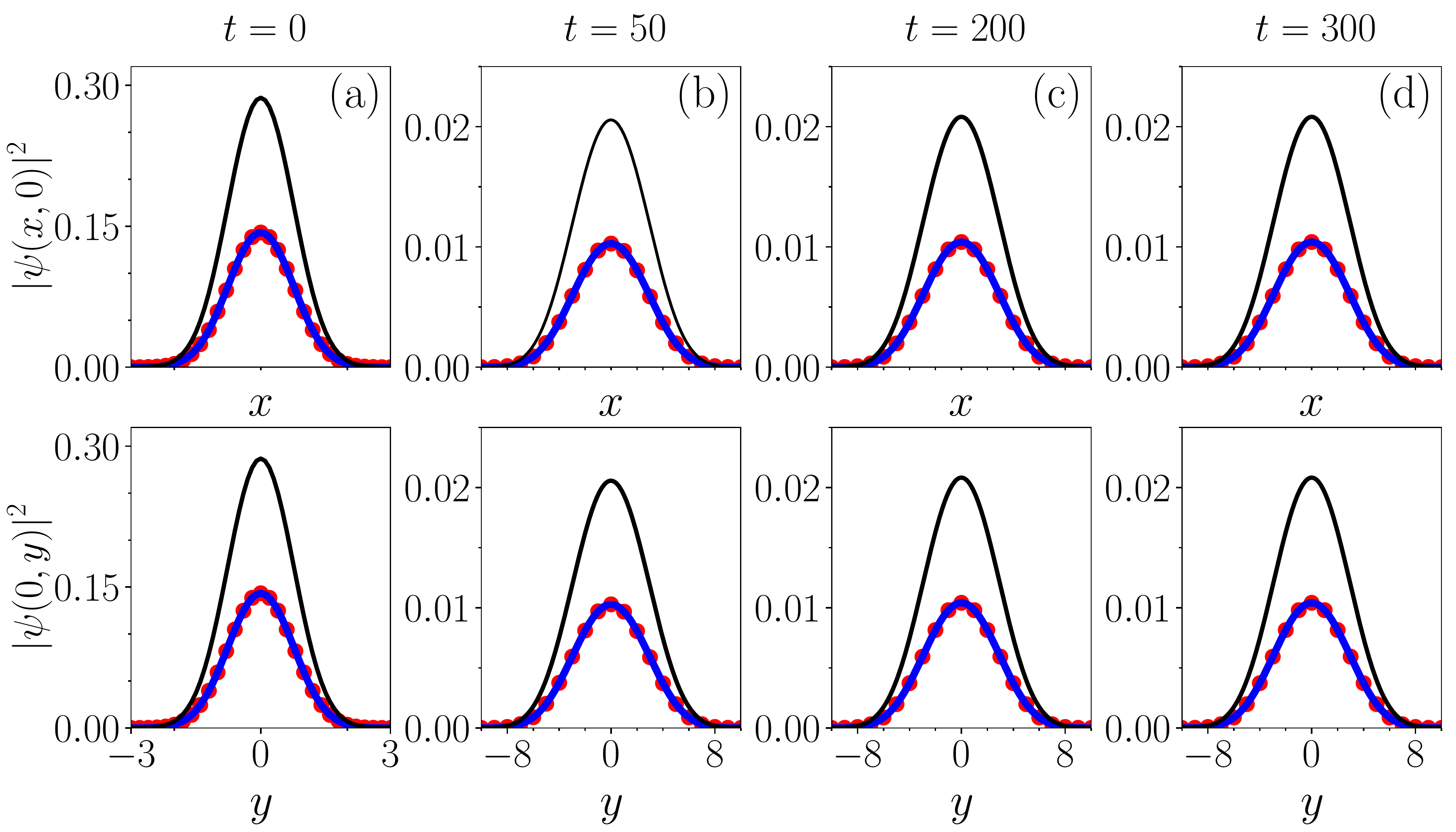}
\caption{One-dimensional density patterns corresponding to Fig.~\ref{fig7}. First row represents the $x$-direction densities where $y=0$, and second row for $y$-direction densities for $x=0$,  $\vert\psi_{\uparrow}\vert^2$ (red line), $\vert\psi_{\downarrow}\vert^2$ (blue line), $\vert\psi_{\uparrow}\vert^2$ + $\vert\psi_{\downarrow}\vert^2$ (black line).}
\label{fig7a}
\end{figure}
To begin with, we consider the parameters $\Omega = 0.75$, $k_L = 0$ and $\alpha = \beta = 1$ with $\lambda = \kappa = 1$ that lie in the stable region. For these parameters we find that the ground state wave-function is of plane wave (PW) nature (See Fig.~\ref{fig7}(a)). In order to analyze the stability of this state we evolved the ground state wave function using RTP. In  Figs.~\ref{fig7}(b-h) we show the temporal evolution of the ground state upto  $t = 300$. As we look at the evolution carefully we find that as time progresses condensate experiences the expansion due to the presence of repulsive interactions. So to stabilize against the expansion trap is added~\cite{Ravisankar2020sol}. At $t=0$ owing to quenching the trap strength by half the condensate expands for a time interval $0<t<50$ (See Fig.~\ref{fig5}(b)). For $t>50$ the size of the condensate  does not change and system attains the stable steady state (Figs.~\ref{fig7}(c-e)). This is evident from the evolution of the density of each spin component as shown in the top and middle rows of  Figs.~\ref{fig7}. Also the similar feature can be seen for the total density (bottom row of Figs.~\ref{fig7}).  Moreover we find that the density profiles of both the components are symmetric in nature and do not exhibit any oscillations (Fig.~\ref{fig7a}). Thus they lack any polarization which is also quite evident from the behaviour of the  eigenvectors that clearly show the presence of the Goldstone mode (in-phase-mode)~\cite{Abad2013, Recati2019} (See first column of Fig~\ref{fig5egvec}).
\begin{figure}[ht]  
\centering \includegraphics[width=0.99\linewidth]{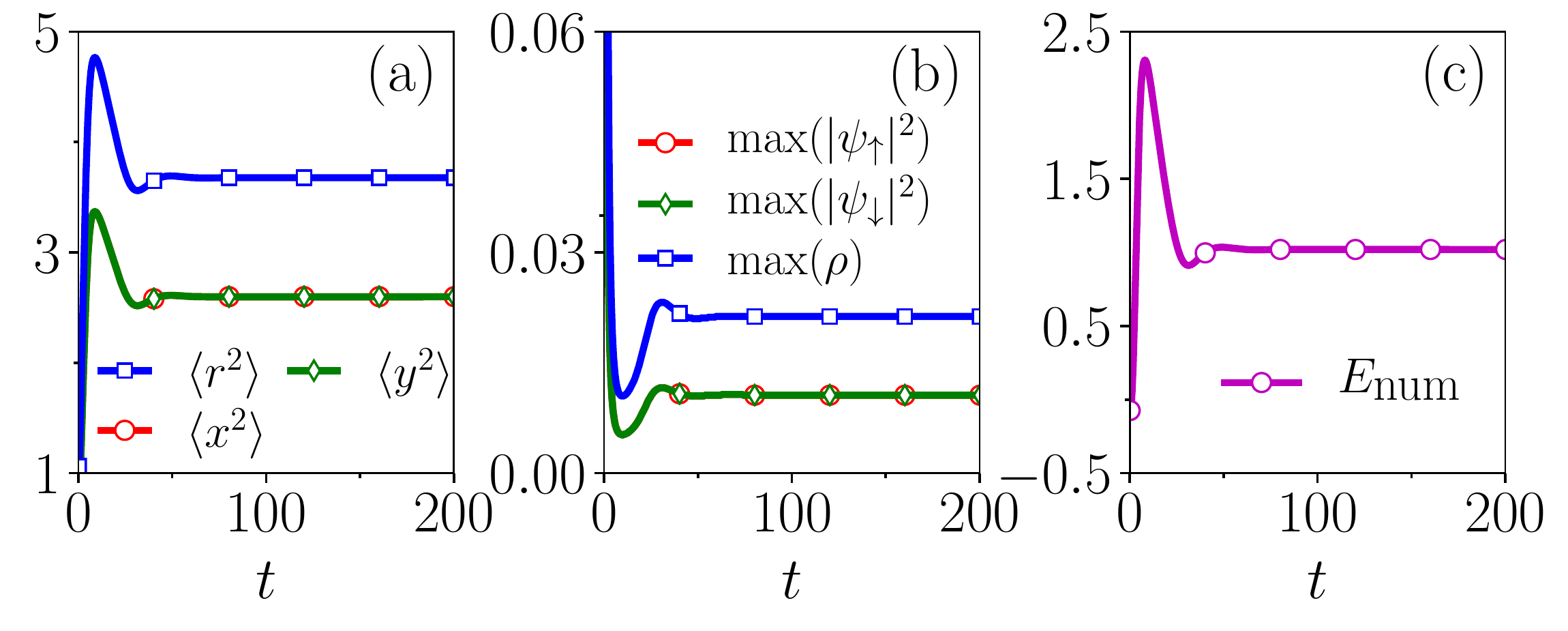} 
\caption{(a) Plot depicts time evolution of root-mean-squared (rms) size in $ \langle x^2 \rangle$ (solid red line), $\langle y^2 \rangle$ (solid green line),  and $\langle r^2 \rangle $ (solid blue line) directions, (b) illustrate the maximum of density $\textrm{max}( \vert \psi_{\uparrow, \downarrow}\vert^2)$ (solid red and green lines), $\textrm{max}(\vert \psi_{\uparrow}\vert^2 + \vert\psi_{\downarrow}\vert^2)$ (solid blue line) and (c) total system energy, the parameters are same as in Fig.~\ref{fig7}.}
\label{fig8}
\end{figure}

In order to verify the dynamics of the condensate next we calculate the root-mean-square (rms) size of the condensate. Initially, we find that the condensate diffuses after the trap is quenched. The rms size of $\langle x^2 \rangle$, $\langle y^2 \rangle$, and $\langle r^2 \rangle$ are illustrated in Fig.~\ref{fig8}(a). The peak densities of both components $\textrm{max}(\vert \psi_{\uparrow, \downarrow}\vert^2)$ and $\textrm{max}(\vert \psi_{\uparrow}\vert^2 + \vert\psi_{\downarrow}\vert^2)$ decrease for a while, thereafter, they attain a steady state which is shown in the Fig.~\ref{fig8}(b). However, the condensate size in both $x$ and $y$ directions is the same as well the maximum density of both components have the similar feature. In Fig.~\ref{fig8}(c) we plot the temporal evolution of the energy which confirms that the system attains the steady state after $t>50$ consistent with the evolution of the density profile. As we analyze the condensate evolution more carefully, as shown in the Figs.~\ref{fig7} and~\ref{fig8}, we find that the ground state is dynamically and energetically stable. It also coincides with the stability regime in the phase diagram of Fig.~\ref{fig6}, which we get from the BdG excitation spectrum.
\begin{figure}
\centering\includegraphics[width=0.99\linewidth]{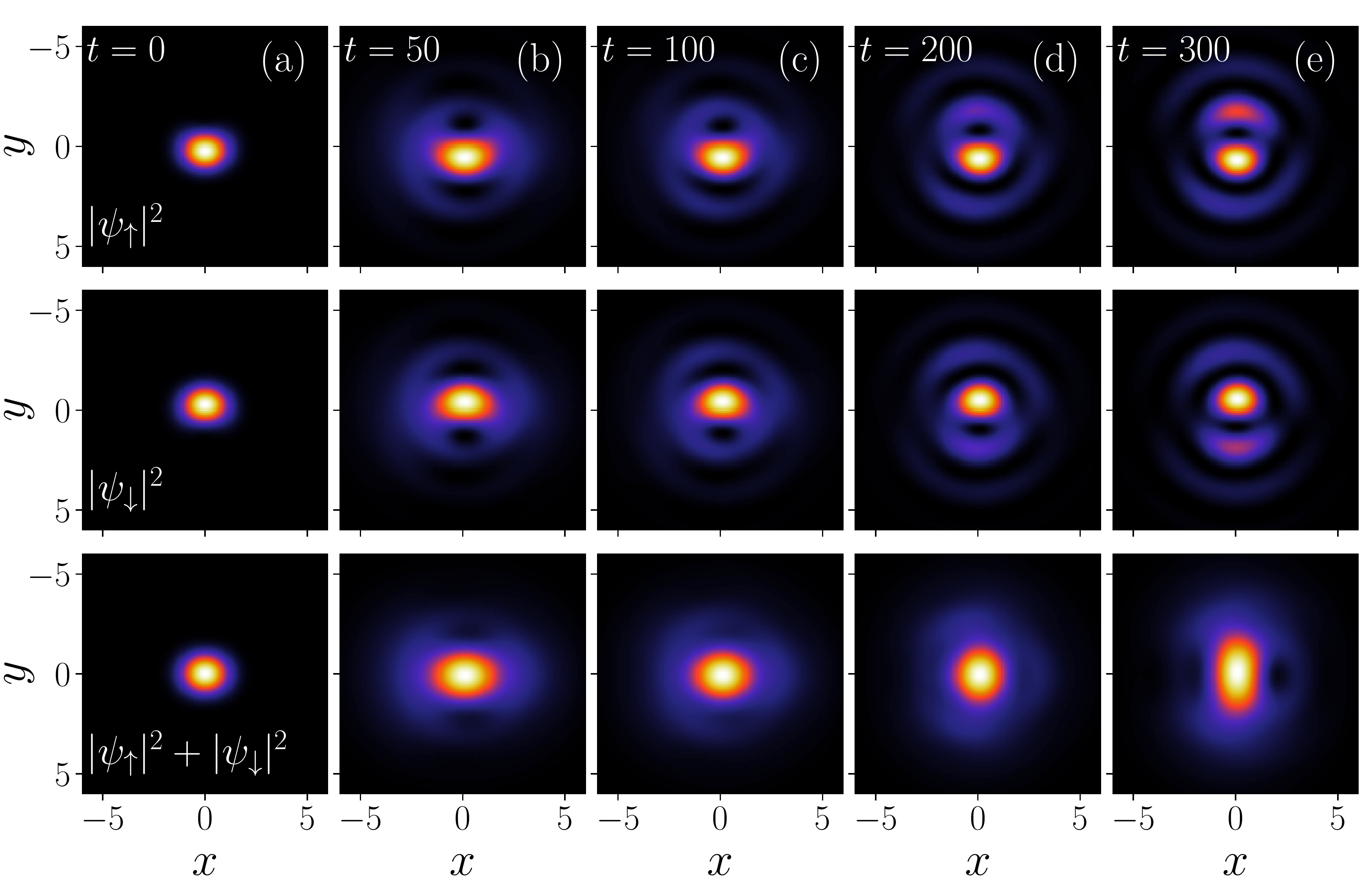}
\caption{Dynamics of ground state density in $x-y$ plane, same as in the Fig.~\ref{fig7} for parameters $k_L = 1.5$, $\Omega = 0.75$ with $\alpha = \beta = 1$. }
\label{fig9a}
\end{figure}
\begin{figure}
\centering\includegraphics[width=0.99\linewidth]{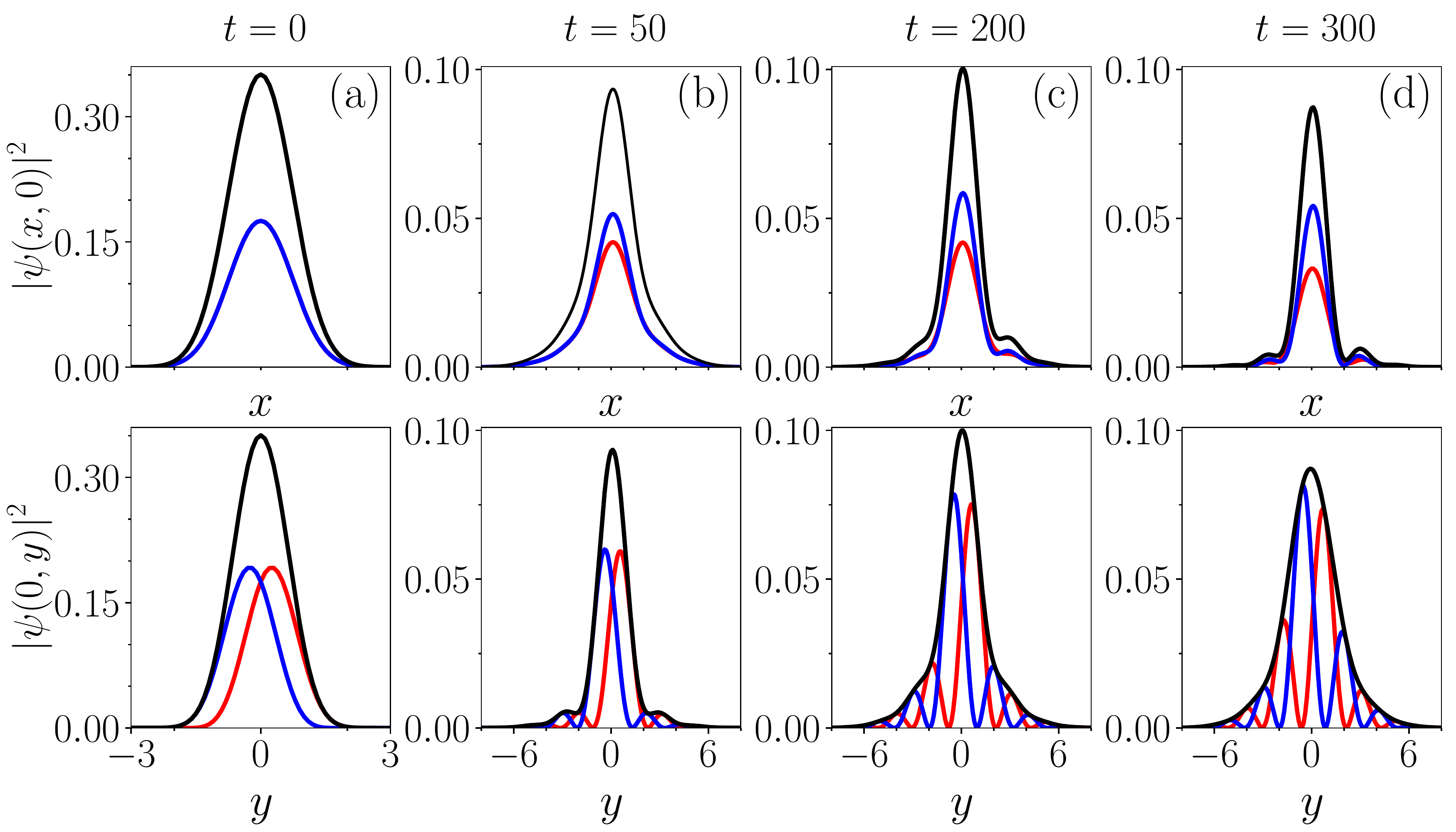}
\caption{One-dimensional density patterns as in Fig.~\ref{fig7a}. The parameters correspond to Fig.~\ref{fig9a}.}
\label{fig9}
\end{figure}

\begin{figure}[!ht]
\centering \includegraphics[width=0.99\linewidth]{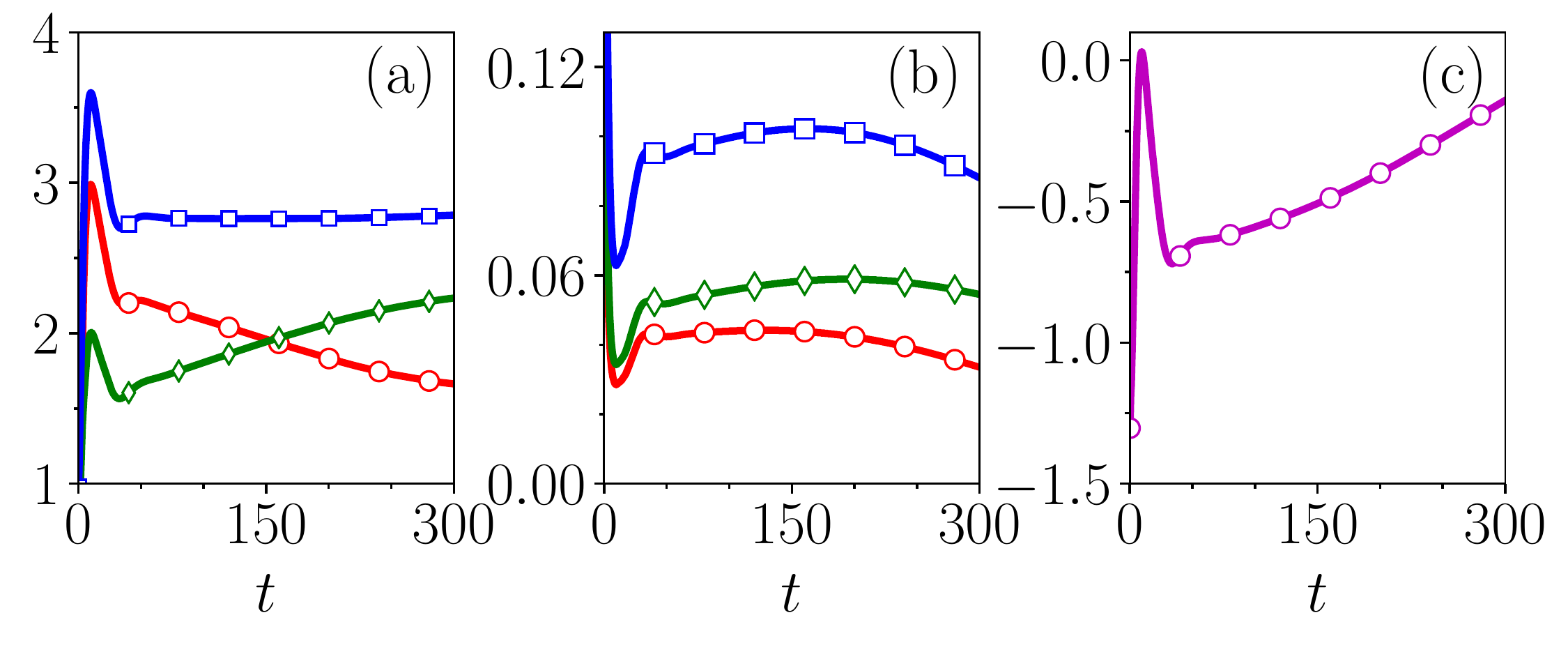} 
\caption{Time evolution of the root-mean-square (rms) size in $\langle x^2 \rangle$ (solid red line), $\langle y^2 \rangle$ (solid green line) and $\langle r^2 \rangle$ (solid blue line) directions, (b) corresponds to the maximum of density $\text{max}(\vert \psi_{\uparrow, \downarrow}\vert^2$ (solid red and green lines), $\textrm{max}(\vert \psi_{\uparrow}\vert^2 + \vert\psi_{\downarrow}\vert^2)$ (solid blue line) and (c) total system energy, the parameters are same as in Fig.~\ref{fig9a}.}
\label{fig10}
\end{figure}

In order to get more insight about the behaviour of the condensate we also analyze the dynamics of instability regime which is investigated by fixing the parameters as $\Omega = 0.75$, $k_L = 1.5$ and $\alpha = \beta = 1$ with trap strength $\lambda = \kappa = 1$. We obtain the ground state as elongated plane wave phase as illustrated in the Fig.~\ref{fig9a}(a).  However, the BdG excitation spectrum have negative energy minimum as well as complex eigen frequencies (see Fig.~\ref{fig5}(d)) in $k_x$  and $k_y$ directions respectively. This feature indicates that the system is energetically (See Fig.~\ref{fig10}(c))  and dynamically (See Fig.~\ref{fig9a}(b-e)) unstable. Apart from this, we also capture the temporal evolution of the the density corresponding the each spin component along with the total density as shown in Fig.~\ref{fig9}. Note that these wave functions are also related to the spin like mode~\cite{Abad2013, Recati2019} as quite evident from the nature of eigenvectors in Fig.~\ref{fig5egvec}(see fourth column).  In Fig.~\ref{fig10} we show the dynamical evolution of different physical quantities like, root-mean-square size of the condensate, maximum of the density and total energy of the system. As we look at the evolution of the condensate rms size in the $x$- and $y$-direction as illustrated in Fig.~\ref{fig10}(a), we  find the condensate size decreases in the $x$-direction while it increases in the $y$-direction with time. However, the spin-components are polarized that leads to the emergence of  interference pattern as time progresses. This suggests the presence of instability in the system which manifestation can also be seen in the temporal evolution of the total density profile as shown in the  bottom row of Fig.~\ref{fig9a}.  Here the density profile that initially was quite symmetric in both directions becomes elongated along the $y$-direction at $t=300$. The other way through which we can characterize the unstable nature of the condensate is by looking at the temporal evolution of the maximum of the the density corresponding to the spin component. Fig.~\ref{fig10}(b) depicts the evolution of the maximum of the spin component as well as maximum corresponding to the total density. After $t\gtrsim 150$ we notice a gradual decrease in their values signifying the unstable behaviour. Finally we show the temporal evolution of the total condensate energy in Fig.~\ref{fig10}(c). We find a sharp increase in the total energy beyond $t\gtrsim 20 $, that clearly indicates towards dynamically unstable nature of the condensate. 

\section{Summary and Conclusions}
\label{sec:7}
In this paper we have investigated the collective excitation spectrum of Rashba SO coupled Bose-Einstein condensate, with Rabi mixing in two-dimensions using the Bogoliubov-de-Gennes theory. First we have analyzed the dispersion of single-particle system in the momentum space, in which as consequences of Rabi coupling the system has broken rotational symmetry, as $k_x$ momentum space has lower ground state energy compared with $k_y$ momentum direction. Following this, we have presented the excitation spectrum with and without nonlinear contact interactions. In the case of noninteracting system, excitations have real negative frequencies, implying that the system is energetically unstable. For interacting case, increases in the Rabi strength for fixed $k_L$ leads the system to make a transition from the unstable phase  to the stable phase. We find  that the phonon-maxon-roton modes disappear upon increase of the Rabi strengths that leads to the loss of the roton gap/depth. The position of roton minimum only present at $k_x = k_L$, and the maxon position approximately equal to the Rabi strengths. The increase of Rabi coupling strength resulted in the loss of instability and bandwidth. However, in contrast to the Rabi coupling, the SO coupling has a destabilizing effect. Further, the Rashba SO coupling turns a symmetric system to an asymmetric one. By increasing the SO strengths, the phonon-maxon-roton modes and instability are revealed in $k_x$ and $k_y$ momentum directions. After a critical value of SO strength, we have negative frequency and additional increase in $k_L$ shows double minimum in the $k_x$ and two instability bands in the $k_y$ momentum direction.

We confirm the observation of the BdG spectrum by directly solving the coupled GPEs. We find the presence of plane wave and elongated plane wave. Their size and maximum of density and energies are studied in the time evolution, which clearly explain about the dynamic and energetic stability and instability phases, also confirms the phenomena from excitation spectrum.

\acknowledgments 

R.R. acknowledges DST-SERB (Department of Science \& Technology - Science and Engineering Research Board) for the financial  support through Project No. ECR/2017/002639 and UGC (University Grants Commission) for financial support in the form of UGC-BSR-RFSMS Research Fellowship scheme (2015-2020). 
A.G. acknowledges FAPESP 2016/17612-7 and CNPq (Conselho Nacional de Desenvolvimento Cient\'{i}fico e Tecnol\'{o}gico) No. 306920/2018-2.
The work of P.M. is supported by CSIR (Council of Scientific and Industrial Research) under Grant No. 03(1422)/18/EMR-II, DST-SERB under Grant No. CRG/2019/004059, FIST (Department of Physics), DST-PURSE and MHRD RUSA 2.0 (Physical Sciences) Programmes, and FAPESP (Funda\c{c}\~{a}o de Amparo \`{a} pesquisa do Estado de S\~{a}o Paulo) Grant No. 2016/00269-8. 
P.K.M. acknowledges  DST-SERB for  the  financial  support  through  Project No. ECR/2017/002639. 

\onecolumngrid
\appendix
\section{A calculation of energy of SO coupled BECs}
\label{app-ene}
In this Appendix we provide the detailed steps to obtain the total energy of the SO coupled BECs. We substitute the stationary state form of the wave function given in  Eq.~(\ref{chem}) in the Eq.~(\ref{eq:gpsoc:1}) and further separate the real and imaginary parts, which yields
\begin{subequations}
\label{eq:gpsoc:reim}
\begin{align}
\mu_{\uparrow} \psi_{\uparrow R} = & \left[ -\frac{1}{2}\frac{\partial^2 }{\partial x^2}  - \frac{1}{2}\frac{\partial^2 }{\partial y^2} +V_{2D}(x,y) +  \alpha \lvert \psi_\uparrow \rvert^2 
+ \beta \lvert \psi_\downarrow \rvert^2\right] \psi _{\uparrow  R} 
+ k_{L} \left(\frac{\partial \psi_{\downarrow I}}{\partial x} - \frac{\partial \psi_{\downarrow R}}{\partial y}  \right) + \Omega  \psi_{\downarrow R}, \label{eq:gpsoc2:reim1}\\
\mu_{\downarrow} \psi_{\downarrow R}  = & \left[ -\frac{1}{2}\frac{\partial^2 }{\partial x^2}  - \frac{1}{2}\frac{\partial^2 }{\partial y^2}+V_{2D}(x,y) + \beta \lvert \psi_\uparrow \rvert^2 
+ \alpha \lvert \psi_\downarrow \rvert^2  \right]\psi _{\downarrow R}
+ k_{L}\left(\frac{\partial \psi_{\uparrow I}}{\partial x} + \frac{\partial \psi_{\uparrow R}}{\partial y} \right) + \Omega \psi_{\uparrow R}, \label{eq:gpsoc2:reim2} 
\end{align}
and for the imaginary parts we have
\begin{align}
\mu_{\uparrow} \psi_{\uparrow I} = & \left[ -\frac{1}{2}\frac{\partial^2 }{\partial x^2}  - \frac{1}{2}\frac{\partial^2 }{\partial y^2} +V_{2D}(x,y) +  \alpha \lvert \psi_\uparrow \rvert^2 
+ \beta \lvert \psi_\downarrow \rvert^2\right] \psi _{\uparrow  I} 
- k_{L} \left(\frac{\partial \psi_{\downarrow R}}{\partial x} + \frac{\partial \psi_{\downarrow I}}{\partial y}  \right) + \Omega  \psi_{\downarrow I}, \label{eq:gpsoc2:reim3}\\
\mu_{\downarrow} \psi_{\downarrow I}  = & \left[ -\frac{1}{2}\frac{\partial^2 }{\partial x^2}  - \frac{1}{2} \frac{\partial^2 }{\partial y^2}+V_{2D}(x,y) + \beta \lvert \psi_\uparrow \rvert^2 
+ \alpha \lvert \psi_\downarrow \rvert^2  \right]\psi _{\downarrow I}
- k_{L}\left(\frac{\partial \psi_{\uparrow R}}{\partial x} - \frac{\partial \psi_{\uparrow I}}{\partial y} \right)  + \Omega \psi_{\uparrow I}, \label{eq:gpsoc2:reim4}
\end{align}
\end{subequations}
where $ \lvert \psi_\uparrow \rvert^2  =   \psi_{\uparrow R}^2 +  \psi_{\uparrow I}^2  $ and  $ \lvert \psi_\downarrow \rvert^2  =   \psi_{\downarrow R}^2 +  \psi_{\downarrow I}^2  $.
Multiplying Eq.~(\ref{eq:gpsoc2:reim1}) with $\psi_{\uparrow R}$ and Eq.~(\ref{eq:gpsoc2:reim2}) with $\psi_{\downarrow R}$, and integrating we get
\begin{subequations}
\begin{align}
\mu_{\uparrow} \int dx\, dy \psi_{\uparrow R}^2 = & \int dx\,dy \psi _{\uparrow  R}\left\{  \left[ -\frac{1}{2}\frac{\partial^2 }{\partial x^2}  - \frac{1}{2}\frac{\partial^2 }{\partial y^2} +V_{2D}(x,y) +  \alpha \lvert \psi_\uparrow \rvert^2 
+ \beta \lvert \psi_\downarrow \rvert^2\right] \psi _{\uparrow  R} 
+   k_{L} \left(\frac{\partial \psi_{\downarrow I}}{\partial x} - \frac{\partial \psi_{\downarrow R}}{\partial y}  \right)
+ \Omega \psi_{\downarrow R}   \right\}, \label{eq:gpsoc2:mu1} \\
\mu_{\downarrow} \int dx\, dy  \psi_{\downarrow R}^2 = & \int dx\,dy \psi _{\downarrow  R} \left\{ \left[ -\frac{1}{2}\frac{\partial^2 }{\partial x^2}  -\frac{1}{2} \frac{\partial^2 }{\partial y^2} +V_{2D}(x,y) +  \beta \lvert \psi_\downarrow \rvert^2 
+ \alpha \lvert \psi_\downarrow \rvert^2\right] \psi _{\downarrow  R} 
-   k_{L} \left(\frac{\partial \psi_{\uparrow I}}{\partial x} + \frac{\partial \psi_{\uparrow R}}{\partial y}  \right)
+ \Omega  \psi_{\uparrow R}   \right\}  , \label{eq:gpsoc2:mu2}
\end{align}
\end{subequations}
Rearranging the above equations yield
\begin{subequations}
\label{eq:gpsoc2:mu:ab}
\begin{align}
\mu_{\uparrow} = & \frac{1}{ \int \psi_{\uparrow R}^2dx\, dy  } \int \left[ \frac{1}{2} \left(\frac{\partial \psi_{\uparrow R}}{\partial x}\right)^2 + \frac{1}{2} \left(\frac{\partial \psi_{\uparrow R}}{\partial y} \right)^2 
+ \left( V_{2D}(x,y) +  \alpha \lvert \psi_\uparrow \rvert^2 + \beta \lvert \psi_\downarrow \rvert^2 \right) \psi_{\uparrow  R}^2  \right]dx\,dy \notag \\
& + \frac{1}{ \int\psi_{\uparrow R}^2 dx\, dy }  \int  \left[ k_{L} \left(\frac{\partial \psi_{\downarrow I}}{\partial x} - \frac{\partial \psi_{\downarrow R}}{\partial y}  \right)
+ \Omega  \psi_{\downarrow R}  \right] \psi_{\uparrow R}  dx\,dy, \label{eq:gpsoc2:mu:1} \\
\mu_{\downarrow} = & \frac{1}{ \int \psi_{\downarrow R}^2dx\, dy  } \int  \left[ \frac{1}{2} \left(\frac{\partial \psi_{\downarrow R}}{\partial x}\right)^2  +\frac{1}{2} \left(\frac{\partial \psi_{\downarrow R}}{\partial y} \right)^2 
+ \left( V_{2D}(x,y) + \beta \lvert \psi_\uparrow \rvert^2  + \alpha \lvert \psi_\uparrow \rvert^2 \right) \psi _{\downarrow  R}^2 \right] dx\,dy \notag \\
& + \frac{1}{ \int\psi_{\downarrow R}^2 dx\, dy }  \int  \left[  k_{L} \left(\frac{\partial \psi_{\uparrow I}}{\partial x} + \frac{\partial \psi_{\uparrow R}}{\partial y}  \right)
+ \Omega  \psi_{\uparrow R}  \right]  \psi_{\downarrow R} dx\,dy, \label{eq:gpsoc2:mu:2}
\end{align}
\end{subequations}
From this we get the total energy is given by $E_{num} = \sum_{j = \uparrow, \downarrow} \frac{\iint \left( E_{j}^{2C} + E_{j}^{SO} \right) \, dx \, dy}  {\iint  \psi_{jR}^2\, dx \, dy}$ can be extracted from the above chemical potential equations as:  
\begin{subequations}%
\begin{align} 
E_{\uparrow}^{2C} &=  \left(\partial_x^2 \psi_{\uparrow R}\right)^2 + \left(\partial_y^2 \psi_{\uparrow R}\right)^2  + \left( V_{2D}  + \alpha \vert \psi_{\uparrow} \vert^2/2  + \beta \vert \psi_{\downarrow} \vert^2/2 \right) \psi_{\uparrow R}^2 \\
E_{\downarrow}^{2C} &=  \left(\partial_x^2 \psi_{\downarrow R}\right)^2 + \left(\partial_y^2 \psi_{\downarrow R}\right)^2  + \left( V_{2D} + \beta \vert \psi_{\uparrow} \vert^2/2  + \alpha \vert \psi_{\downarrow} \vert^2/2 \right) \psi_{\downarrow R}^2 \\
E_{\uparrow}^{SO} &=  k_{L} \left(\psi_{\uparrow R} \partial_{x} \psi_{\downarrow I} - \psi_{\uparrow R} \partial_{y} \psi_{\downarrow R} \right) + \psi_{\uparrow R} \Omega \psi_{\downarrow R} \\
E_{\downarrow}^{SO} &=  k_{L} \left(\psi_{\downarrow R} \partial_{x} \psi_{\uparrow I} + \psi_{\downarrow R} \partial_{y} \psi_{\uparrow R} \right) + \psi_{\downarrow R} \Omega \psi_{\uparrow R}
\end{align}
\end{subequations}%

\section{Elements of the BdG matrix}
\label{app-matrix}
\begin{subequations}%
\begin{align}
f(n_{\uparrow},n_{\downarrow}) & =  \frac{\left( k^2_x + k^2_y \right)}{2} + 2 \alpha n_{\uparrow} +\beta n_{\downarrow} -\frac{1}{2} \left[ \alpha (n_{\uparrow} + n_{\downarrow})+\beta n - \frac{n \Omega}{\sqrt{n_{\uparrow} n_{\downarrow}}} \right], \\ 
g(n_{\uparrow},n_{\downarrow}) & =  \frac{\left(k^2_x + k^2_y \right)}{2} + \alpha n_{\uparrow} + 2 \beta n_{\downarrow} -\frac{1}{2} \left[ \beta (n_{\uparrow} + n_{\downarrow}) + \alpha n - \frac{n \Omega}{\sqrt{n_{\uparrow} n_{\downarrow}}} \right], \\ 
L_{13(24)}  &= \beta \sqrt{n_{\uparrow} n_{\downarrow}} \pm k_L \left( k_x - \mathrm{i} k_y \right)-\Omega,  \\ 
L_{31(42)} & = \beta \sqrt{n_{\uparrow} n_{\downarrow}} \pm k_L \left(k_x + \mathrm{i} k_y \right)-\Omega,  
\end{align}
\end{subequations}%

\section{Coefficients of the BdG excitation spectrum}
\label{app-coeff}
\begin{subequations}%
\begin{align}
b = & - \left(k_x^2 + k_y^2\right) \left[ 2 \left(k_L ^2 + \Omega \right) \alpha + \frac{1}{2} \left( k_x^2 + k_y^2 \right) \right] - 2 \Omega \left[(\alpha - \beta) +2 \Omega \right], \\
c = & 2 k_L k_x \Big[2 \Omega  \left[ (\alpha - \beta) + 2 \Omega \right] - \left(k_x^2 + k_y^2\right) (\beta - 2 \Omega )\Big],\\
d = & \frac{1}{16}\bigg[\left(k_x^8 + k_y^8 \right) + 4P \left(k_x^6 + k_y^6 \right) + 2Q \left(k_x^4 + k_y^4\right) + k_x^2 k_y^2\left[\left(k_x^4 + k_y^4\right) + 12P \left(k_x^2 + k_y^2 \right) + 6 k_x^2 k_y^2 + 4Q \right] \bigg] 
+ \left(R k_x^2 + S k_y^2\right) 
\end{align}
\end{subequations}%
where
\begin{subequations}%
\begin{align}
P & = \alpha -2 k_L ^2+2 \Omega \\
Q & = 2\left[\alpha^2-\beta ^2+\alpha  \left(6 \Omega -4 k_L ^2\right)+2 \beta  \Omega + 4 \left(k_L ^2-\Omega \right)^2\right] \\
R & = \Omega  \left(\alpha +\beta - 2 k_L ^2\right) (\alpha - \beta + 2 \Omega ) \\
S & = \Omega  (\alpha +\beta ) \left(\alpha -\beta -2 k_L ^2+2 \Omega \right) 
\end{align}
\end{subequations}%

\twocolumngrid


\begin{thebibliography}{48}%
\makeatletter
\providecommand \@ifxundefined [1]{%
 \@ifx{#1\undefined}
}%
\providecommand \@ifnum [1]{%
 \ifnum #1\expandafter \@firstoftwo
 \else \expandafter \@secondoftwo
 \fi
}%
\providecommand \@ifx [1]{%
 \ifx #1\expandafter \@firstoftwo
 \else \expandafter \@secondoftwo
 \fi
}%
\providecommand \natexlab [1]{#1}%
\providecommand \enquote  [1]{``#1''}%
\providecommand \bibnamefont  [1]{#1}%
\providecommand \bibfnamefont [1]{#1}%
\providecommand \citenamefont [1]{#1}%
\providecommand \href@noop [0]{\@secondoftwo}%
\providecommand \href [0]{\begingroup \@sanitize@url \@href}%
\providecommand \@href[1]{\@@startlink{#1}\@@href}%
\providecommand \@@href[1]{\endgroup#1\@@endlink}%
\providecommand \@sanitize@url [0]{\catcode `\\12\catcode `\$12\catcode
  `\&12\catcode `\#12\catcode `\^12\catcode `\_12\catcode `\%12\relax}%
\providecommand \@@startlink[1]{}%
\providecommand \@@endlink[0]{}%
\providecommand \url  [0]{\begingroup\@sanitize@url \@url }%
\providecommand \@url [1]{\endgroup\@href {#1}{\urlprefix }}%
\providecommand \urlprefix  [0]{URL }%
\providecommand \Eprint [0]{\href }%
\providecommand \doibase [0]{http://dx.doi.org/}%
\providecommand \selectlanguage [0]{\@gobble}%
\providecommand \bibinfo  [0]{\@secondoftwo}%
\providecommand \bibfield  [0]{\@secondoftwo}%
\providecommand \translation [1]{[#1]}%
\providecommand \BibitemOpen [0]{}%
\providecommand \bibitemStop [0]{}%
\providecommand \bibitemNoStop [0]{.\EOS\space}%
\providecommand \EOS [0]{\spacefactor3000\relax}%
\providecommand \BibitemShut  [1]{\csname bibitem#1\endcsname}%
\let\auto@bib@innerbib\@empty
\bibitem [{\citenamefont {Anderson}\ \emph {et~al.}(1995)\citenamefont
  {Anderson}, \citenamefont {Ensher}, \citenamefont {Matthews}, \citenamefont
  {Wieman},\ and\ \citenamefont {Cornell}}]{Anderson1995}%
  \BibitemOpen
  \bibfield  {author} {\bibinfo {author} {\bibfnamefont {M.~H.}\ \bibnamefont
  {Anderson}}, \bibinfo {author} {\bibfnamefont {J.~R.}\ \bibnamefont
  {Ensher}}, \bibinfo {author} {\bibfnamefont {M.~R.}\ \bibnamefont
  {Matthews}}, \bibinfo {author} {\bibfnamefont {C.~E.}\ \bibnamefont
  {Wieman}}, \ and\ \bibinfo {author} {\bibfnamefont {E.~A.}\ \bibnamefont
  {Cornell}},\ }\href {\doibase 10.1126/science.269.5221.198} {\bibfield
  {journal} {\bibinfo  {journal} {Science}\ }\textbf {\bibinfo {volume}
  {269}},\ \bibinfo {pages} {198} (\bibinfo {year} {1995})}\BibitemShut
  {NoStop}%
\bibitem [{\citenamefont {Davis}\ \emph {et~al.}(1995)\citenamefont {Davis},
  \citenamefont {Mewes}, \citenamefont {Andrews}, \citenamefont {van Druten},
  \citenamefont {Durfee}, \citenamefont {Kurn},\ and\ \citenamefont
  {Ketterle}}]{Davis1995}%
  \BibitemOpen
  \bibfield  {author} {\bibinfo {author} {\bibfnamefont {K.~B.}\ \bibnamefont
  {Davis}}, \bibinfo {author} {\bibfnamefont {M.~O.}\ \bibnamefont {Mewes}},
  \bibinfo {author} {\bibfnamefont {M.~R.}\ \bibnamefont {Andrews}}, \bibinfo
  {author} {\bibfnamefont {N.~J.}\ \bibnamefont {van Druten}}, \bibinfo
  {author} {\bibfnamefont {D.~S.}\ \bibnamefont {Durfee}}, \bibinfo {author}
  {\bibfnamefont {D.~M.}\ \bibnamefont {Kurn}}, \ and\ \bibinfo {author}
  {\bibfnamefont {W.}~\bibnamefont {Ketterle}},\ }\href {\doibase
  10.1103/physrevlett.75.3969} {\bibfield  {journal} {\bibinfo  {journal}
  {Phys. Rev. Lett.}\ }\textbf {\bibinfo {volume} {75}},\ \bibinfo {pages}
  {3969} (\bibinfo {year} {1995})}\BibitemShut {NoStop}%
\bibitem [{\citenamefont {Bradley}\ \emph {et~al.}(1995)\citenamefont
  {Bradley}, \citenamefont {Sackett}, \citenamefont {Tollett},\ and\
  \citenamefont {Hulet}}]{Bradley1995}%
  \BibitemOpen
  \bibfield  {author} {\bibinfo {author} {\bibfnamefont {C.~C.}\ \bibnamefont
  {Bradley}}, \bibinfo {author} {\bibfnamefont {C.~A.}\ \bibnamefont
  {Sackett}}, \bibinfo {author} {\bibfnamefont {J.~J.}\ \bibnamefont
  {Tollett}}, \ and\ \bibinfo {author} {\bibfnamefont {R.~G.}\ \bibnamefont
  {Hulet}},\ }\href {\doibase 10.1103/physrevlett.75.1687} {\bibfield
  {journal} {\bibinfo  {journal} {Phys. Rev. Lett.}\ }\textbf {\bibinfo
  {volume} {75}},\ \bibinfo {pages} {1687} (\bibinfo {year}
  {1995})}\BibitemShut {NoStop}%
\bibitem [{\citenamefont {Gerton}\ \emph {et~al.}(2000)\citenamefont {Gerton},
  \citenamefont {Strekalov}, \citenamefont {Prodan},\ and\ \citenamefont
  {Hulet}}]{Gerton2000}%
  \BibitemOpen
  \bibfield  {author} {\bibinfo {author} {\bibfnamefont {J.~M.}\ \bibnamefont
  {Gerton}}, \bibinfo {author} {\bibfnamefont {D.}~\bibnamefont {Strekalov}},
  \bibinfo {author} {\bibfnamefont {I.}~\bibnamefont {Prodan}}, \ and\ \bibinfo
  {author} {\bibfnamefont {R.~G.}\ \bibnamefont {Hulet}},\ }\href {\doibase
  10.1038/35047030} {\bibfield  {journal} {\bibinfo  {journal} {Nature}\
  }\textbf {\bibinfo {volume} {408}},\ \bibinfo {pages} {692} (\bibinfo {year}
  {2000})}\BibitemShut {NoStop}%
\bibitem [{\citenamefont {Greiner}\ \emph {et~al.}(2002)\citenamefont
  {Greiner}, \citenamefont {Mandel}, \citenamefont {Esslinger}, \citenamefont
  {Hänsch},\ and\ \citenamefont {Bloch}}]{Greiner2002}%
  \BibitemOpen
  \bibfield  {author} {\bibinfo {author} {\bibfnamefont {M.}~\bibnamefont
  {Greiner}}, \bibinfo {author} {\bibfnamefont {O.}~\bibnamefont {Mandel}},
  \bibinfo {author} {\bibfnamefont {T.}~\bibnamefont {Esslinger}}, \bibinfo
  {author} {\bibfnamefont {T.~W.}\ \bibnamefont {Hänsch}}, \ and\ \bibinfo
  {author} {\bibfnamefont {I.}~\bibnamefont {Bloch}},\ }\href {\doibase
  10.1038/415039a} {\bibfield  {journal} {\bibinfo  {journal} {Nature}\
  }\textbf {\bibinfo {volume} {415}},\ \bibinfo {pages} {39} (\bibinfo {year}
  {2002})}\BibitemShut {NoStop}%
\bibitem [{\citenamefont {Morsch}\ and\ \citenamefont
  {Oberthaler}(2006)}]{Morsch2006}%
  \BibitemOpen
  \bibfield  {author} {\bibinfo {author} {\bibfnamefont {O.}~\bibnamefont
  {Morsch}}\ and\ \bibinfo {author} {\bibfnamefont {M.}~\bibnamefont
  {Oberthaler}},\ }\href {\doibase 10.1103/revmodphys.78.179} {\bibfield
  {journal} {\bibinfo  {journal} {Rev. Mod. Phys.}\ }\textbf {\bibinfo {volume}
  {78}},\ \bibinfo {pages} {179} (\bibinfo {year} {2006})}\BibitemShut
  {NoStop}%
\bibitem [{\citenamefont {Lewenstein}\ \emph {et~al.}(2007)\citenamefont
  {Lewenstein}, \citenamefont {Sanpera}, \citenamefont {Ahufinger},
  \citenamefont {Damski}, \citenamefont {Sen(De)},\ and\ \citenamefont
  {Sen}}]{Lewenstein2007}%
  \BibitemOpen
  \bibfield  {author} {\bibinfo {author} {\bibfnamefont {M.}~\bibnamefont
  {Lewenstein}}, \bibinfo {author} {\bibfnamefont {A.}~\bibnamefont {Sanpera}},
  \bibinfo {author} {\bibfnamefont {V.}~\bibnamefont {Ahufinger}}, \bibinfo
  {author} {\bibfnamefont {B.}~\bibnamefont {Damski}}, \bibinfo {author}
  {\bibfnamefont {A.}~\bibnamefont {Sen(De)}}, \ and\ \bibinfo {author}
  {\bibfnamefont {U.}~\bibnamefont {Sen}},\ }\href {\doibase
  10.1080/00018730701223200} {\bibfield  {journal} {\bibinfo  {journal} {Adv.
  Phys.}\ }\textbf {\bibinfo {volume} {56}},\ \bibinfo {pages} {243} (\bibinfo
  {year} {2007})}\BibitemShut {NoStop}%
\bibitem [{\citenamefont {Roati}\ \emph {et~al.}(2008)\citenamefont {Roati},
  \citenamefont {D'Errico}, \citenamefont {Fallani}, \citenamefont {Fattori},
  \citenamefont {Fort}, \citenamefont {Zaccanti}, \citenamefont {Modugno},
  \citenamefont {Modugno},\ and\ \citenamefont {Inguscio}}]{Roati2008}%
  \BibitemOpen
  \bibfield  {author} {\bibinfo {author} {\bibfnamefont {G.}~\bibnamefont
  {Roati}}, \bibinfo {author} {\bibfnamefont {C.}~\bibnamefont {D'Errico}},
  \bibinfo {author} {\bibfnamefont {L.}~\bibnamefont {Fallani}}, \bibinfo
  {author} {\bibfnamefont {M.}~\bibnamefont {Fattori}}, \bibinfo {author}
  {\bibfnamefont {C.}~\bibnamefont {Fort}}, \bibinfo {author} {\bibfnamefont
  {M.}~\bibnamefont {Zaccanti}}, \bibinfo {author} {\bibfnamefont
  {G.}~\bibnamefont {Modugno}}, \bibinfo {author} {\bibfnamefont
  {M.}~\bibnamefont {Modugno}}, \ and\ \bibinfo {author} {\bibfnamefont
  {M.}~\bibnamefont {Inguscio}},\ }\href {\doibase 10.1038/nature07071}
  {\bibfield  {journal} {\bibinfo  {journal} {Nature}\ }\textbf {\bibinfo
  {volume} {453}},\ \bibinfo {pages} {895} (\bibinfo {year}
  {2008})}\BibitemShut {NoStop}%
\bibitem [{\citenamefont {Chin}\ \emph {et~al.}(2010)\citenamefont {Chin},
  \citenamefont {Grimm}, \citenamefont {Julienne},\ and\ \citenamefont
  {Tiesinga}}]{Chin2010}%
  \BibitemOpen
  \bibfield  {author} {\bibinfo {author} {\bibfnamefont {C.}~\bibnamefont
  {Chin}}, \bibinfo {author} {\bibfnamefont {R.}~\bibnamefont {Grimm}},
  \bibinfo {author} {\bibfnamefont {P.}~\bibnamefont {Julienne}}, \ and\
  \bibinfo {author} {\bibfnamefont {E.}~\bibnamefont {Tiesinga}},\ }\href
  {\doibase 10.1103/revmodphys.82.1225} {\bibfield  {journal} {\bibinfo
  {journal} {Rev. Mod. Phys.}\ }\textbf {\bibinfo {volume} {82}},\ \bibinfo
  {pages} {1225} (\bibinfo {year} {2010})}\BibitemShut {NoStop}%
\bibitem [{\citenamefont {Griesmaier}\ \emph {et~al.}(2005)\citenamefont
  {Griesmaier}, \citenamefont {Werner}, \citenamefont {Hensler}, \citenamefont
  {Stuhler},\ and\ \citenamefont {Pfau}}]{Griesmaier2005}%
  \BibitemOpen
  \bibfield  {author} {\bibinfo {author} {\bibfnamefont {A.}~\bibnamefont
  {Griesmaier}}, \bibinfo {author} {\bibfnamefont {J.}~\bibnamefont {Werner}},
  \bibinfo {author} {\bibfnamefont {S.}~\bibnamefont {Hensler}}, \bibinfo
  {author} {\bibfnamefont {J.}~\bibnamefont {Stuhler}}, \ and\ \bibinfo
  {author} {\bibfnamefont {T.}~\bibnamefont {Pfau}},\ }\href {\doibase
  10.1103/physrevlett.94.160401} {\bibfield  {journal} {\bibinfo  {journal}
  {Phys. Rev. Lett.}\ }\textbf {\bibinfo {volume} {94}} (\bibinfo {year}
  {2005})}\BibitemShut {NoStop}%
\bibitem [{\citenamefont {Lu}\ \emph {et~al.}(2011)\citenamefont {Lu},
  \citenamefont {Burdick}, \citenamefont {Youn},\ and\ \citenamefont
  {Lev}}]{Lu2011}%
  \BibitemOpen
  \bibfield  {author} {\bibinfo {author} {\bibfnamefont {M.}~\bibnamefont
  {Lu}}, \bibinfo {author} {\bibfnamefont {N.~Q.}\ \bibnamefont {Burdick}},
  \bibinfo {author} {\bibfnamefont {S.~H.}\ \bibnamefont {Youn}}, \ and\
  \bibinfo {author} {\bibfnamefont {B.~L.}\ \bibnamefont {Lev}},\ }\href
  {\doibase 10.1103/physrevlett.107.190401} {\bibfield  {journal} {\bibinfo
  {journal} {Phys. Rev. Lett.}\ }\textbf {\bibinfo {volume} {107}} (\bibinfo
  {year} {2011})}\BibitemShut {NoStop}%
\bibitem [{\citenamefont {Aikawa}\ \emph {et~al.}(2012)\citenamefont {Aikawa},
  \citenamefont {Frisch}, \citenamefont {Mark}, \citenamefont {Baier},
  \citenamefont {Rietzler}, \citenamefont {Grimm},\ and\ \citenamefont
  {Ferlaino}}]{Aikawa2012}%
  \BibitemOpen
  \bibfield  {author} {\bibinfo {author} {\bibfnamefont {K.}~\bibnamefont
  {Aikawa}}, \bibinfo {author} {\bibfnamefont {A.}~\bibnamefont {Frisch}},
  \bibinfo {author} {\bibfnamefont {M.}~\bibnamefont {Mark}}, \bibinfo {author}
  {\bibfnamefont {S.}~\bibnamefont {Baier}}, \bibinfo {author} {\bibfnamefont
  {A.}~\bibnamefont {Rietzler}}, \bibinfo {author} {\bibfnamefont
  {R.}~\bibnamefont {Grimm}}, \ and\ \bibinfo {author} {\bibfnamefont
  {F.}~\bibnamefont {Ferlaino}},\ }\href {\doibase
  10.1103/physrevlett.108.210401} {\bibfield  {journal} {\bibinfo  {journal}
  {Phys. Rev. Lett.}\ }\textbf {\bibinfo {volume} {108}},\ \bibinfo {pages}
  {210401} (\bibinfo {year} {2012})}\BibitemShut {NoStop}%
\bibitem [{\citenamefont {Lin}\ \emph {et~al.}(2011)\citenamefont {Lin},
  \citenamefont {Jim{\'{e}}nez-Garc{\'{\i}}a},\ and\ \citenamefont
  {Spielman}}]{Lin2011}%
  \BibitemOpen
  \bibfield  {author} {\bibinfo {author} {\bibfnamefont {Y.-J.}\ \bibnamefont
  {Lin}}, \bibinfo {author} {\bibfnamefont {K.}~\bibnamefont
  {Jim{\'{e}}nez-Garc{\'{\i}}a}}, \ and\ \bibinfo {author} {\bibfnamefont
  {I.~B.}\ \bibnamefont {Spielman}},\ }\href {\doibase 10.1038/nature09887}
  {\bibfield  {journal} {\bibinfo  {journal} {Nature}\ }\textbf {\bibinfo
  {volume} {471}},\ \bibinfo {pages} {83} (\bibinfo {year} {2011})}\BibitemShut
  {NoStop}%
\bibitem [{\citenamefont {Galitski}\ and\ \citenamefont
  {Spielman}(2013)}]{Galitski2013}%
  \BibitemOpen
  \bibfield  {author} {\bibinfo {author} {\bibfnamefont {V.}~\bibnamefont
  {Galitski}}\ and\ \bibinfo {author} {\bibfnamefont {I.~B.}\ \bibnamefont
  {Spielman}},\ }\href {\doibase 10.1038/nature11841} {\bibfield  {journal}
  {\bibinfo  {journal} {Nature}\ }\textbf {\bibinfo {volume} {494}},\ \bibinfo
  {pages} {49} (\bibinfo {year} {2013})}\BibitemShut {NoStop}%
\bibitem [{\citenamefont {Ho}(1998)}]{Ho1998}%
  \BibitemOpen
  \bibfield  {author} {\bibinfo {author} {\bibfnamefont {T.-L.}\ \bibnamefont
  {Ho}},\ }\href {\doibase 10.1103/physrevlett.81.742} {\bibfield  {journal}
  {\bibinfo  {journal} {Phys. Rev. Lett.}\ }\textbf {\bibinfo {volume} {81}},\
  \bibinfo {pages} {742} (\bibinfo {year} {1998})}\BibitemShut {NoStop}%
\bibitem [{\citenamefont {Papp}\ \emph {et~al.}(2008)\citenamefont {Papp},
  \citenamefont {Pino},\ and\ \citenamefont {Wieman}}]{Papp2008}%
  \BibitemOpen
  \bibfield  {author} {\bibinfo {author} {\bibfnamefont {S.~B.}\ \bibnamefont
  {Papp}}, \bibinfo {author} {\bibfnamefont {J.~M.}\ \bibnamefont {Pino}}, \
  and\ \bibinfo {author} {\bibfnamefont {C.~E.}\ \bibnamefont {Wieman}},\
  }\href {\doibase 10.1103/physrevlett.101.040402} {\bibfield  {journal}
  {\bibinfo  {journal} {Phys. Rev. Lett.}\ }\textbf {\bibinfo {volume} {101}},\
  \bibinfo {pages} {040402} (\bibinfo {year} {2008})}\BibitemShut {NoStop}%
\bibitem [{\citenamefont {Takeuchi}\ \emph {et~al.}(2010)\citenamefont
  {Takeuchi}, \citenamefont {Ishino},\ and\ \citenamefont
  {Tsubota}}]{Takeuchi2010}%
  \BibitemOpen
  \bibfield  {author} {\bibinfo {author} {\bibfnamefont {H.}~\bibnamefont
  {Takeuchi}}, \bibinfo {author} {\bibfnamefont {S.}~\bibnamefont {Ishino}}, \
  and\ \bibinfo {author} {\bibfnamefont {M.}~\bibnamefont {Tsubota}},\ }\href
  {\doibase 10.1103/physrevlett.105.205301} {\bibfield  {journal} {\bibinfo
  {journal} {Phys. Rev. Lett.}\ }\textbf {\bibinfo {volume} {105}},\ \bibinfo
  {pages} {205301} (\bibinfo {year} {2010})}\BibitemShut {NoStop}%
\bibitem [{\citenamefont {Sabbatini}\ \emph {et~al.}(2011)\citenamefont
  {Sabbatini}, \citenamefont {Zurek},\ and\ \citenamefont
  {Davis}}]{Sabbatini2011}%
  \BibitemOpen
  \bibfield  {author} {\bibinfo {author} {\bibfnamefont {J.}~\bibnamefont
  {Sabbatini}}, \bibinfo {author} {\bibfnamefont {W.~H.}\ \bibnamefont
  {Zurek}}, \ and\ \bibinfo {author} {\bibfnamefont {M.~J.}\ \bibnamefont
  {Davis}},\ }\href {\doibase 10.1103/physrevlett.107.230402} {\bibfield
  {journal} {\bibinfo  {journal} {Phys. Rev. Lett.}\ }\textbf {\bibinfo
  {volume} {107}},\ \bibinfo {pages} {230402} (\bibinfo {year}
  {2011})}\BibitemShut {NoStop}%
\bibitem [{\citenamefont {Law}\ \emph {et~al.}(2010)\citenamefont {Law},
  \citenamefont {Kevrekidis},\ and\ \citenamefont {Tuckerman}}]{Law2010}%
  \BibitemOpen
  \bibfield  {author} {\bibinfo {author} {\bibfnamefont {K.~J.~H.}\
  \bibnamefont {Law}}, \bibinfo {author} {\bibfnamefont {P.~G.}\ \bibnamefont
  {Kevrekidis}}, \ and\ \bibinfo {author} {\bibfnamefont {L.~S.}\ \bibnamefont
  {Tuckerman}},\ }\href {\doibase 10.1103/physrevlett.105.160405} {\bibfield
  {journal} {\bibinfo  {journal} {Phys. Rev. Lett.}\ }\textbf {\bibinfo
  {volume} {105}},\ \bibinfo {pages} {160405} (\bibinfo {year}
  {2010})}\BibitemShut {NoStop}%
\bibitem [{\citenamefont {Wen}\ \emph {et~al.}(2013)\citenamefont {Wen},
  \citenamefont {Qiao}, \citenamefont {Xu},\ and\ \citenamefont
  {Mao}}]{Wen2013}%
  \BibitemOpen
  \bibfield  {author} {\bibinfo {author} {\bibfnamefont {L.}~\bibnamefont
  {Wen}}, \bibinfo {author} {\bibfnamefont {Y.}~\bibnamefont {Qiao}}, \bibinfo
  {author} {\bibfnamefont {Y.}~\bibnamefont {Xu}}, \ and\ \bibinfo {author}
  {\bibfnamefont {L.}~\bibnamefont {Mao}},\ }\href {\doibase
  10.1103/physreva.87.033604} {\bibfield  {journal} {\bibinfo  {journal} {Phys.
  Rev. A}\ }\textbf {\bibinfo {volume} {87}},\ \bibinfo {pages} {033604}
  (\bibinfo {year} {2013})}\BibitemShut {NoStop}%
\bibitem [{\citenamefont {Aidelsburger}\ \emph {et~al.}(2013)\citenamefont
  {Aidelsburger}, \citenamefont {Atala}, \citenamefont {Lohse}, \citenamefont
  {Barreiro}, \citenamefont {Paredes},\ and\ \citenamefont
  {Bloch}}]{Aidelsburger2013}%
  \BibitemOpen
  \bibfield  {author} {\bibinfo {author} {\bibfnamefont {M.}~\bibnamefont
  {Aidelsburger}}, \bibinfo {author} {\bibfnamefont {M.}~\bibnamefont {Atala}},
  \bibinfo {author} {\bibfnamefont {M.}~\bibnamefont {Lohse}}, \bibinfo
  {author} {\bibfnamefont {J.~T.}\ \bibnamefont {Barreiro}}, \bibinfo {author}
  {\bibfnamefont {B.}~\bibnamefont {Paredes}}, \ and\ \bibinfo {author}
  {\bibfnamefont {I.}~\bibnamefont {Bloch}},\ }\href {\doibase
  10.1103/physrevlett.111.185301} {\bibfield  {journal} {\bibinfo  {journal}
  {Phys. Rev. Lett.}\ }\textbf {\bibinfo {volume} {111}},\ \bibinfo {pages}
  {185301} (\bibinfo {year} {2013})}\BibitemShut {NoStop}%
\bibitem [{\citenamefont {Goldman}\ \emph {et~al.}(2010)\citenamefont
  {Goldman}, \citenamefont {Satija}, \citenamefont {Nikolic}, \citenamefont
  {Bermudez}, \citenamefont {Martin-Delgado}, \citenamefont {Lewenstein},\ and\
  \citenamefont {Spielman}}]{Goldman2010}%
  \BibitemOpen
  \bibfield  {author} {\bibinfo {author} {\bibfnamefont {N.}~\bibnamefont
  {Goldman}}, \bibinfo {author} {\bibfnamefont {I.}~\bibnamefont {Satija}},
  \bibinfo {author} {\bibfnamefont {P.}~\bibnamefont {Nikolic}}, \bibinfo
  {author} {\bibfnamefont {A.}~\bibnamefont {Bermudez}}, \bibinfo {author}
  {\bibfnamefont {M.~A.}\ \bibnamefont {Martin-Delgado}}, \bibinfo {author}
  {\bibfnamefont {M.}~\bibnamefont {Lewenstein}}, \ and\ \bibinfo {author}
  {\bibfnamefont {I.~B.}\ \bibnamefont {Spielman}},\ }\href {\doibase
  10.1103/PhysRevLett.105.255302} {\bibfield  {journal} {\bibinfo  {journal}
  {Phys. Rev. Lett.}\ }\textbf {\bibinfo {volume} {105}},\ \bibinfo {pages}
  {255302} (\bibinfo {year} {2010})}\BibitemShut {NoStop}%
\bibitem [{\citenamefont {Wu}\ \emph {et~al.}(2016)\citenamefont {Wu},
  \citenamefont {Zhang}, \citenamefont {Sun}, \citenamefont {Xu}, \citenamefont
  {Wang}, \citenamefont {Ji}, \citenamefont {Deng}, \citenamefont {Chen},
  \citenamefont {Liu},\ and\ \citenamefont {Pan}}]{Wu2016}%
  \BibitemOpen
  \bibfield  {author} {\bibinfo {author} {\bibfnamefont {Z.}~\bibnamefont
  {Wu}}, \bibinfo {author} {\bibfnamefont {L.}~\bibnamefont {Zhang}}, \bibinfo
  {author} {\bibfnamefont {W.}~\bibnamefont {Sun}}, \bibinfo {author}
  {\bibfnamefont {X.-T.}\ \bibnamefont {Xu}}, \bibinfo {author} {\bibfnamefont
  {B.-Z.}\ \bibnamefont {Wang}}, \bibinfo {author} {\bibfnamefont {S.-C.}\
  \bibnamefont {Ji}}, \bibinfo {author} {\bibfnamefont {Y.}~\bibnamefont
  {Deng}}, \bibinfo {author} {\bibfnamefont {S.}~\bibnamefont {Chen}}, \bibinfo
  {author} {\bibfnamefont {X.-J.}\ \bibnamefont {Liu}}, \ and\ \bibinfo
  {author} {\bibfnamefont {J.-W.}\ \bibnamefont {Pan}},\ }\href {\doibase
  10.1126/science.aaf6689} {\bibfield  {journal} {\bibinfo  {journal}
  {Science}\ }\textbf {\bibinfo {volume} {354}},\ \bibinfo {pages} {83}
  (\bibinfo {year} {2016})}\BibitemShut {NoStop}%
\bibitem [{\citenamefont {Seaman}\ \emph {et~al.}(2007)\citenamefont {Seaman},
  \citenamefont {Kr\"amer}, \citenamefont {Anderson},\ and\ \citenamefont
  {Holland}}]{Seaman2007}%
  \BibitemOpen
  \bibfield  {author} {\bibinfo {author} {\bibfnamefont {B.~T.}\ \bibnamefont
  {Seaman}}, \bibinfo {author} {\bibfnamefont {M.}~\bibnamefont {Kr\"amer}},
  \bibinfo {author} {\bibfnamefont {D.~Z.}\ \bibnamefont {Anderson}}, \ and\
  \bibinfo {author} {\bibfnamefont {M.~J.}\ \bibnamefont {Holland}},\ }\href
  {\doibase 10.1103/PhysRevA.75.023615} {\bibfield  {journal} {\bibinfo
  {journal} {Phys. Rev. A}\ }\textbf {\bibinfo {volume} {75}},\ \bibinfo
  {pages} {023615} (\bibinfo {year} {2007})}\BibitemShut {NoStop}%
\bibitem [{\citenamefont {Andrianov}\ and\ \citenamefont
  {Moiseev}(2014)}]{Andrianov2014}%
  \BibitemOpen
  \bibfield  {author} {\bibinfo {author} {\bibfnamefont {S.~N.}\ \bibnamefont
  {Andrianov}}\ and\ \bibinfo {author} {\bibfnamefont {S.~A.}\ \bibnamefont
  {Moiseev}},\ }\href {\doibase 10.1103/physreva.90.042303} {\bibfield
  {journal} {\bibinfo  {journal} {Phys. Rev. A}\ }\textbf {\bibinfo {volume}
  {90}},\ \bibinfo {pages} {042303} (\bibinfo {year} {2014})}\BibitemShut
  {NoStop}%
\bibitem{Ravisankar2020sol}
R. Ravisankar, T. Sriraman, L. Salasnich, and P. Muruganandam,
J. Phys. B {\bf 53}, 195301 (2020).
\bibitem{Achilleos2013-bs}
V. Achilleos, D. J. Frantzeskakis, P. G. Kevrekidis, and D. E. Pelinovsky,
Phys. Rev. Lett. {\bf 110}, 264101 (2013).
\bibitem [{\citenamefont {Achilleos}\ \emph {et~al.}(2013)\citenamefont
  {Achilleos}, \citenamefont {Stockhofe}, \citenamefont {Kevrekidis},
  \citenamefont {Frantzeskakis},\ and\ \citenamefont
  {Schmelcher}}]{Achilleos2013}%
  \BibitemOpen
  \bibfield  {author} {\bibinfo {author} {\bibfnamefont {V.}~\bibnamefont
  {Achilleos}}, \bibinfo {author} {\bibfnamefont {J.}~\bibnamefont
  {Stockhofe}}, \bibinfo {author} {\bibfnamefont {P.~G.}\ \bibnamefont
  {Kevrekidis}}, \bibinfo {author} {\bibfnamefont {D.~J.}\ \bibnamefont
  {Frantzeskakis}}, \ and\ \bibinfo {author} {\bibfnamefont {P.}~\bibnamefont
  {Schmelcher}},\ }\href {\doibase 10.1209/0295-5075/103/20002} {\bibfield
  {journal} {\bibinfo  {journal} {{Europhys. Lett.}}\ }\textbf {\bibinfo {volume} {103}},\
  \bibinfo {pages} {20002} (\bibinfo {year} {2013})}\BibitemShut {NoStop}%
\bibitem [{\citenamefont {Jin}\ \emph {et~al.}(2014)\citenamefont {Jin},
  \citenamefont {Zhang},\ and\ \citenamefont {Han}}]{Jin2014}%
  \BibitemOpen
  \bibfield  {author} {\bibinfo {author} {\bibfnamefont {J.}~\bibnamefont
  {Jin}}, \bibinfo {author} {\bibfnamefont {S.}~\bibnamefont {Zhang}}, \ and\
  \bibinfo {author} {\bibfnamefont {W.}~\bibnamefont {Han}},\ }\href {\doibase
  10.1088/0953-4075/47/11/115302} {\bibfield  {journal} {\bibinfo  {journal}
  {J. Phys. B}\ }\textbf {\bibinfo {volume} {47}},\ \bibinfo {pages} {115302}
  (\bibinfo {year} {2014})}\BibitemShut {NoStop}%
\bibitem [{\citenamefont {Li}\ and\ \citenamefont {Sakaguchi}(2013)}]{Li2013}%
  \BibitemOpen
  \bibfield  {author} {\bibinfo {author} {\bibfnamefont {B.}~\bibnamefont
  {Li}}\ and\ \bibinfo {author} {\bibfnamefont {H.}~\bibnamefont {Sakaguchi}},\
  }\href {\doibase 10.1007/s10909-013-0988-1} {\bibfield  {journal} {\bibinfo
  {journal} {J. Low. Temp. Phys.}\ }\textbf {\bibinfo {volume} {175}},\
  \bibinfo {pages} {243} (\bibinfo {year} {2013})}\BibitemShut {NoStop}%
\bibitem [{\citenamefont {Cheng}\ \emph {et~al.}(2014)\citenamefont {Cheng},
  \citenamefont {Tang},\ and\ \citenamefont {Adhikari}}]{Cheng2014}%
  \BibitemOpen
  \bibfield  {author} {\bibinfo {author} {\bibfnamefont {Y.}~\bibnamefont
  {Cheng}}, \bibinfo {author} {\bibfnamefont {G.}~\bibnamefont {Tang}}, \ and\
  \bibinfo {author} {\bibfnamefont {S.~K.}\ \bibnamefont {Adhikari}},\ }\href
  {\doibase 10.1103/PhysRevA.89.063602} {\bibfield  {journal} {\bibinfo
  {journal} {Phys. Rev. A}\ }\textbf {\bibinfo {volume} {89}},\ \bibinfo
  {pages} {063602} (\bibinfo {year} {2014})}\BibitemShut {NoStop}%
\bibitem{He2021}
H. He and Y. Zhang,
Phys. Rev. A {\bf 103}, 053322 (2021).
\bibitem{Pal2017}
S. Pal, A. Roy, and D. Angom,
J. Phys. B: At. Mol. Opt. Phys. {\bf 50}, 195301 (2017).
\bibitem{Pal2018}
S. Pal, A. Roy, and D. Angom,
J. Phys. B: At. Mol. Opt. Phys. {\bf 51}, 085302 (2018).
\bibitem [{\citenamefont {Martone}\ \emph {et~al.}(2012)\citenamefont
  {Martone}, \citenamefont {Li}, \citenamefont {Pitaevskii},\ and\
  \citenamefont {Stringari}}]{Martone2012}%
  \BibitemOpen
  \bibfield  {author} {\bibinfo {author} {\bibfnamefont {G.~I.}\ \bibnamefont
  {Martone}}, \bibinfo {author} {\bibfnamefont {Y.}~\bibnamefont {Li}},
  \bibinfo {author} {\bibfnamefont {L.~P.}\ \bibnamefont {Pitaevskii}}, \ and\
  \bibinfo {author} {\bibfnamefont {S.}~\bibnamefont {Stringari}},\ }\href
  {\doibase 10.1103/PhysRevA.86.063621} {\bibfield  {journal} {\bibinfo
  {journal} {Phys. Rev. A}\ }\textbf {\bibinfo {volume} {86}},\ \bibinfo
  {pages} {063621} (\bibinfo {year} {2012})}\BibitemShut {NoStop}%
\bibitem [{\citenamefont {Zheng}\ \emph {et~al.}(2013)\citenamefont {Zheng},
  \citenamefont {Yu}, \citenamefont {Cui},\ and\ \citenamefont
  {Zhai}}]{Zheng2013}%
  \BibitemOpen
  \bibfield  {author} {\bibinfo {author} {\bibfnamefont {W.}~\bibnamefont
  {Zheng}}, \bibinfo {author} {\bibfnamefont {Z.-Q.}\ \bibnamefont {Yu}},
  \bibinfo {author} {\bibfnamefont {X.}~\bibnamefont {Cui}}, \ and\ \bibinfo
  {author} {\bibfnamefont {H.}~\bibnamefont {Zhai}},\ }\href {\doibase
  10.1088/0953-4075/46/13/134007} {\bibfield  {journal} {\bibinfo  {journal}
  {J. Phys. B}\ }\textbf {\bibinfo {volume} {46}},\ \bibinfo {pages} {134007}
  (\bibinfo {year} {2013})}\BibitemShut {NoStop}%
\bibitem [{\citenamefont {Khamehchi}\ \emph {et~al.}(2014)\citenamefont
  {Khamehchi}, \citenamefont {Zhang}, \citenamefont {Hamner}, \citenamefont
  {Busch},\ and\ \citenamefont {Engels}}]{Khamehchi2014}%
  \BibitemOpen
  \bibfield  {author} {\bibinfo {author} {\bibfnamefont {M.~A.}\ \bibnamefont
  {Khamehchi}}, \bibinfo {author} {\bibfnamefont {Y.}~\bibnamefont {Zhang}},
  \bibinfo {author} {\bibfnamefont {C.}~\bibnamefont {Hamner}}, \bibinfo
  {author} {\bibfnamefont {T.}~\bibnamefont {Busch}}, \ and\ \bibinfo {author}
  {\bibfnamefont {P.}~\bibnamefont {Engels}},\ }\href {\doibase
  10.1103/PhysRevA.90.063624} {\bibfield  {journal} {\bibinfo  {journal} {Phys.
  Rev. A}\ }\textbf {\bibinfo {volume} {90}},\ \bibinfo {pages} {063624}
  (\bibinfo {year} {2014})}\BibitemShut {NoStop}%
\bibitem [{\citenamefont {Ji}\ \emph {et~al.}(2015)\citenamefont {Ji},
  \citenamefont {Zhang}, \citenamefont {Xu}, \citenamefont {Wu}, \citenamefont
  {Deng}, \citenamefont {Chen},\ and\ \citenamefont {Pan}}]{Ji2015}%
  \BibitemOpen
  \bibfield  {author} {\bibinfo {author} {\bibfnamefont {S.-C.}\ \bibnamefont
  {Ji}}, \bibinfo {author} {\bibfnamefont {L.}~\bibnamefont {Zhang}}, \bibinfo
  {author} {\bibfnamefont {X.-T.}\ \bibnamefont {Xu}}, \bibinfo {author}
  {\bibfnamefont {Z.}~\bibnamefont {Wu}}, \bibinfo {author} {\bibfnamefont
  {Y.}~\bibnamefont {Deng}}, \bibinfo {author} {\bibfnamefont {S.}~\bibnamefont
  {Chen}}, \ and\ \bibinfo {author} {\bibfnamefont {J.-W.}\ \bibnamefont
  {Pan}},\ }\href {\doibase 10.1103/PhysRevLett.114.105301} {\bibfield
  {journal} {\bibinfo  {journal} {Phys. Rev. Lett.}\ }\textbf {\bibinfo
  {volume} {114}},\ \bibinfo {pages} {105301} (\bibinfo {year}
  {2015})}\BibitemShut {NoStop}%
\bibitem [{\citenamefont {Liao}\ \emph {et~al.}(2015)\citenamefont {Liao},
  \citenamefont {Fialko}, \citenamefont {Brand},\ and\ \citenamefont
  {Zülicke}}]{Liao2015}%
  \BibitemOpen
  \bibfield  {author} {\bibinfo {author} {\bibfnamefont {R.}~\bibnamefont
  {Liao}}, \bibinfo {author} {\bibfnamefont {O.}~\bibnamefont {Fialko}},
  \bibinfo {author} {\bibfnamefont {J.}~\bibnamefont {Brand}}, \ and\ \bibinfo
  {author} {\bibfnamefont {U.}~\bibnamefont {Zülicke}},\ }\href {\doibase
  10.1103/physreva.92.043633} {\bibfield  {journal} {\bibinfo  {journal} {Phys.
  Rev. A}\ }\textbf {\bibinfo {volume} {92}},\ \bibinfo {pages} {043633}
  (\bibinfo {year} {2015})}\BibitemShut {NoStop}%
\bibitem{Ozawa2013}
T. Ozawa, L. P. Pitaevskii, and S. Stringari,
Phys. Rev. A {\bf 87}, 063610 (2013).
\bibitem{Chen2017}
L. Chen, H. Pu, Z-Q. Yu, and Y. Zhang,
Phys. Rev. A {\bf 95}, 033616 (2017).
\bibitem{Yu2018}
Z-F. Yu and J-K. Xue,
Europhys. Lett. 121, 20003 (2018).
\bibitem{Geier2021}
K. T. Geier, G. I. Martone, P. Hauke, and S. Stringari,
arXiv:2102.02221v1 (2021).
\bibitem{Zhu2012}
Q. Zhu1, C. Zhang, and B. Wu,
Europhys. Lett. {\bf 100}, 50003 (2012).
\bibitem [{\citenamefont {Pu}\ and\ \citenamefont {Bigelow}(1998)}]{Pu1998}%
  \BibitemOpen
  \bibfield  {author} {\bibinfo {author} {\bibfnamefont {H.}~\bibnamefont
  {Pu}}\ and\ \bibinfo {author} {\bibfnamefont {N.~P.}\ \bibnamefont
  {Bigelow}},\ }\href {\doibase 10.1103/physrevlett.80.1130} {\bibfield
  {journal} {\bibinfo  {journal} {Phys. Rev. Lett.}\ }\textbf {\bibinfo
  {volume} {80}},\ \bibinfo {pages} {1130} (\bibinfo {year}
  {1998})}\BibitemShut {NoStop}%
\bibitem [{\citenamefont {Yu}(2013)}]{Yu2013}%
  \BibitemOpen
  \bibfield  {author} {\bibinfo {author} {\bibfnamefont {Z.-Q.}\ \bibnamefont
  {Yu}},\ }\href {\doibase 10.1103/PhysRevA.87.051606} {\bibfield  {journal}
  {\bibinfo  {journal} {Phys. Rev. A}\ }\textbf {\bibinfo {volume} {87}},\
  \bibinfo {pages} {051606} (\bibinfo {year} {2013})}\BibitemShut {NoStop}%
\bibitem{Sahu2020}
S. Sahu and D. Majumder, 
J. Phys.B: At. Mol. Opt. Phys. 53, 095301 (2020).
\bibitem [{\citenamefont {Wang}\ \emph {et~al.}(2010)\citenamefont {Wang},
  \citenamefont {Gao}, \citenamefont {Jian},\ and\ \citenamefont
  {Zhai}}]{Wang2010}%
  \BibitemOpen
  \bibfield  {author} {\bibinfo {author} {\bibfnamefont {C.}~\bibnamefont
  {Wang}}, \bibinfo {author} {\bibfnamefont {C.}~\bibnamefont {Gao}}, \bibinfo
  {author} {\bibfnamefont {C.-M.}\ \bibnamefont {Jian}}, \ and\ \bibinfo
  {author} {\bibfnamefont {H.}~\bibnamefont {Zhai}},\ }\href {\doibase
  10.1103/PhysRevLett.105.160403} {\bibfield  {journal} {\bibinfo  {journal}
  {Phys. Rev. Lett.}\ }\textbf {\bibinfo {volume} {105}},\ \bibinfo {pages}
  {160403} (\bibinfo {year} {2010})}\BibitemShut {NoStop}%
\bibitem [{\citenamefont {Goldstein}\ and\ \citenamefont
  {Meystre}(1997)}]{Goldstein1997}%
  \BibitemOpen
  \bibfield  {author} {\bibinfo {author} {\bibfnamefont {E.~V.}\ \bibnamefont
  {Goldstein}}\ and\ \bibinfo {author} {\bibfnamefont {P.}~\bibnamefont
  {Meystre}},\ }\href {\doibase 10.1103/physreva.55.2935} {\bibfield  {journal}
  {\bibinfo  {journal} {Phys. Rev. A}\ }\textbf {\bibinfo {volume} {55}},\
  \bibinfo {pages} {2935} (\bibinfo {year} {1997})}\BibitemShut {NoStop}%
\bibitem [{\citenamefont {Abad}\ and\ \citenamefont {Recati}(2013)}]{Abad2013}%
  \BibitemOpen
  \bibfield  {author} {\bibinfo {author} {\bibfnamefont {M.}~\bibnamefont
  {Abad}}\ and\ \bibinfo {author} {\bibfnamefont {A.}~\bibnamefont {Recati}},\
  }\href {\doibase 10.1140/epjd/e2013-40053-2} {\bibfield  {journal} {\bibinfo
  {journal} {Eur. Phys. J. D}\ }\textbf {\bibinfo {volume} {67}},\ \bibinfo
  {pages} {148} (\bibinfo {year} {2013})}\BibitemShut {NoStop}%
\bibitem{Tommasini2003}
P. Tommasini, E. J. V. de Passos, A. F. R. de Toledo Piza, M. S.Hussein, and E. Timmermans
Phys. Rev. A { \bf 67}, 023606 (2003).
\bibitem{Recati2019}
A. Recati and F. Piazza,
Phys. Rev. B {\bf 99}, 064505 (2019).
\bibitem{Li2012} 
Y. Li, L. P. Pitaevskii, and S. Stringari, 
Phys. Rev. Lett.  {\bf 108}, 225301 (2012).
\bibitem{Ravisankar2020}
R. Ravisankar, T. Sriraman, and P. Muruganandam,
AIP Conf. Proc. {\bf 2265}, 030022 (2020).
\bibitem{Bhuvaneswari2018}
S. Bhuvaneswari, K. Nithyanandan, and P. Muruganandam,
J. Phys. Commun. {\bf 2}, 025008 (2018).
\bibitem [{\citenamefont {Bogolyubov}(1947)}]{Bogolyubov1947}%
  \BibitemOpen
  \bibfield  {author} {\bibinfo {author} {\bibfnamefont {N.~N.}\ \bibnamefont
  {Bogolyubov}},\ }\href@noop {} {\bibfield  {journal} {\bibinfo  {journal} {J.
  Phys. (USSR)}\ }\textbf {\bibinfo {volume} {11}},\ \bibinfo {pages} {23}
  (\bibinfo {year} {1947})}\BibitemShut {NoStop}%
\bibitem [{\citenamefont {Zilsel}(1950)}]{Zilsel1950}%
  \BibitemOpen
  \bibfield  {author} {\bibinfo {author} {\bibfnamefont {P.~R.}\ \bibnamefont
  {Zilsel}},\ }\href {\doibase 10.1103/PhysRev.79.309} {\bibfield  {journal}
  {\bibinfo  {journal} {Phys. Rev.}\ }\textbf {\bibinfo {volume} {79}},\
  \bibinfo {pages} {309} (\bibinfo {year} {1950})}\BibitemShut {NoStop}%
\bibitem [{\citenamefont {Rickayzen}(1959)}]{Rickayzen1959}%
  \BibitemOpen
  \bibfield  {author} {\bibinfo {author} {\bibfnamefont {G.}~\bibnamefont
  {Rickayzen}},\ }\href {\doibase 10.1103/PhysRev.115.795} {\bibfield
  {journal} {\bibinfo  {journal} {Phys. Rev.}\ }\textbf {\bibinfo {volume}
  {115}},\ \bibinfo {pages} {795} (\bibinfo {year} {1959})}\BibitemShut
  {NoStop}%
\bibitem [{\citenamefont {Jin}\ \emph {et~al.}(1996)\citenamefont {Jin},
  \citenamefont {Ensher}, \citenamefont {Matthews}, \citenamefont {Wieman},\
  and\ \citenamefont {Cornell}}]{Jin1996}%
  \BibitemOpen
  \bibfield  {author} {\bibinfo {author} {\bibfnamefont {D.~S.}\ \bibnamefont
  {Jin}}, \bibinfo {author} {\bibfnamefont {J.~R.}\ \bibnamefont {Ensher}},
  \bibinfo {author} {\bibfnamefont {M.~R.}\ \bibnamefont {Matthews}}, \bibinfo
  {author} {\bibfnamefont {C.~E.}\ \bibnamefont {Wieman}}, \ and\ \bibinfo
  {author} {\bibfnamefont {E.~A.}\ \bibnamefont {Cornell}},\ }\href {\doibase
  10.1103/PhysRevLett.77.420} {\bibfield  {journal} {\bibinfo  {journal} {Phys.
  Rev. Lett.}\ }\textbf {\bibinfo {volume} {77}},\ \bibinfo {pages} {420}
  (\bibinfo {year} {1996})}\BibitemShut {NoStop}%
\bibitem [{\citenamefont {Mewes}\ \emph {et~al.}(1996)\citenamefont {Mewes},
  \citenamefont {Andrews}, \citenamefont {van Druten}, \citenamefont {Kurn},
  \citenamefont {Durfee}, \citenamefont {Townsend},\ and\ \citenamefont
  {Ketterle}}]{Mewes1996}%
  \BibitemOpen
  \bibfield  {author} {\bibinfo {author} {\bibfnamefont {M.-O.}\ \bibnamefont
  {Mewes}}, \bibinfo {author} {\bibfnamefont {M.~R.}\ \bibnamefont {Andrews}},
  \bibinfo {author} {\bibfnamefont {N.~J.}\ \bibnamefont {van Druten}},
  \bibinfo {author} {\bibfnamefont {D.~M.}\ \bibnamefont {Kurn}}, \bibinfo
  {author} {\bibfnamefont {D.~S.}\ \bibnamefont {Durfee}}, \bibinfo {author}
  {\bibfnamefont {C.~G.}\ \bibnamefont {Townsend}}, \ and\ \bibinfo {author}
  {\bibfnamefont {W.}~\bibnamefont {Ketterle}},\ }\href {\doibase
  10.1103/PhysRevLett.77.988} {\bibfield  {journal} {\bibinfo  {journal} {Phys.
  Rev. Lett.}\ }\textbf {\bibinfo {volume} {77}},\ \bibinfo {pages} {988}
  (\bibinfo {year} {1996})}\BibitemShut {NoStop}%
\bibitem{Anderson1999}
E. Anderson, Z. Bai, C. Bischof, S. Blackford, J. Demmel, J.  Dongarra,J. Du Croz, A. Greenbaum, S. Hammarling, A. McKenney, D. Sorensen, 
{\it{{LAPACK} Users' Guide}},
Society for Industrial and Applied Mathematics, Third Ed (1999).


\bibitem [{\citenamefont {Muruganandam}\ and\ \citenamefont
  {Adhikari}(2009)}]{Muruganandam2009}%
  \BibitemOpen
  \bibfield  {author} {\bibinfo {author} {\bibfnamefont {P.}~\bibnamefont
  {Muruganandam}}\ and\ \bibinfo {author} {\bibfnamefont {S.~K.}\ \bibnamefont
  {Adhikari}},\ }\href {\doibase 10.1016/j.cpc.2009.04.015} {\bibfield
  {journal} {\bibinfo  {journal} {Comput. Phys. Commun.}\ }\textbf {\bibinfo
  {volume} {180}},\ \bibinfo {pages} {1888} (\bibinfo {year}
  {2009})}\BibitemShut {NoStop}%
\bibitem{Young2016} 
L. E. Young-S., D. Vudragovi\'{c}, P. Muruganandam, S. K. Adhikari, and A. Bala\v{z},
Comput. Phys. Commun. {\bf 204}, 209 (2016).
\bibitem{Ravisankar2021}
R Ravisankar, D Vudragovi\'{c}, P Muruganandam, A Bala\v{z}, SK Adhikari,
Comput. Phys. Commun. {\bf 259}, 107657 (2021).
\bibitem{Muruganandam2021}
P. Muruganandam, S. K. Adhikari, and A. Bala\v{z},
Comput. Phys. Commun. {\bf 264}, 107926 (2021).
\end{thebibliography}


%

\end{document}